\documentclass[final,3p,times,twocolumn]{elsarticle}
\usepackage{graphicx}
\usepackage{amssymb}
\usepackage[utf8]{inputenc}
\usepackage{textcomp}

\usepackage{booktabs}
\usepackage{rotating}
\usepackage{multirow}

\usepackage{url,hyperref}

\usepackage{color}
\usepackage[usenames,dvipsnames,svgnames,table]{xcolor}

\usepackage{caption}
\usepackage{subcaption}

\journal{Journal of Nuclear Materials }

\begin{document}

\begin{frontmatter}



\title{Simulation of the nanostructure evolution under irradiation in Fe-C
alloys}


\author[sck,helsinki]{V.~Jansson\corref{cor1}}
\ead{ville.b.c.jansson@gmail.com}
\author[sck]{L.~Malerba}

\cortext[cor1]{Corresponding author. Tel. +32 1433 3096, fax: +32 1432 1216.}

\address[sck]{Institute of Nuclear Materials Science, SCK$\bullet$CEN, Boeretang
200, 2400
{\sc Mol, Belgium}}

\address[helsinki]{Department of Physics, P.O. Box 43 (Pehr
Kalms gata 2), FI-00014 {\sc University of Helsinki, Finland}}

\begin{abstract}

Neutron irradiation induces in steels nanostructural changes, which are at the
origin of the mechanical degradation that these materials experience during
operation in nuclear power plants. Some of these effects can be studied by using
as model alloy the iron-carbon system.

The Object Kinetic Monte Carlo technique has proven capable of simulating in a
realistic and quantitatively reliable way a whole irradiation process. We have developed a 
model for simulating Fe-C systems using a physical description of the properties
of vacancy and self-interstitial
atom (SIA) clusters, based on a selection of the latest data from atomistic
studies and other available experimental
and theoretical work from the literature. Based on these data, the effect of
carbon on radiation defect evolution has been largely understood in terms of formation of immobile complexes with vacancies that in turn act as traps for SIA clusters. It is found
that this effect can be introduced using generic traps for SIA and vacancy clusters, with
a binding
energy that depends on the size of the clusters, also chosen on the basis on
previously performed atomistic studies. 

The model proved suitable to reproduce the results of low ($<$350 K) 
temperature neutron irradiation experiments, as well as the
corresponding post-irradiation annealing up to 700 K, in terms of defect cluster
densities and size distribution, when compared to available experimental data
from the literature. The use of traps proved instrumental for our model.
\end{abstract}


\end{frontmatter}

\frenchspacing

% \linenumbers

\section{Introduction}

\noindent Low alloy banitic steels are used for the reactor pressure vessel
(RPV) in most
commercial nuclear power plants. Their integrity is the
most important factor in the determination of the lifetime of a nuclear reactor.
It is known that neutron irradiation causes hardening and reduction of the
tensile elongation of these steels, leading to an increase of the
ductile-brittle transition temperature (embrittlement)
\cite{little1976neutron,odette1998recent,odette2001embrittlement,
zinkle2006microstructure}.
For a full understanding of the degradation processes of steels, it is important
to understand thoroughly how the nanostructure evolves under irradiation,
\textit{i.e.} how vacancies, SIA and their clusters
interact
with each other and with other impurities, such as interstitial carbon atoms.

Object Kinetic Monte Carlo (OKMC) is a stochastic method that is adequate for
simulating the nanostructure evolution in materials during irradiation or
annealing. However, the method requires the pre-determination of the parameters
that define the rate of events, such as migration of defects or dissociation
of clusters, for all defects in the simulation. Since there are many kinds of
defects, the properties of which usually vary with the size of the defect
clusters, thousands of parameters are needed for a standard simulation. These may be obtained by using data from experiments, \textit{ab initio}
calculations and MD or other atomistic calculations. OKMC
studies have been performed for Fe-systems \textit{e.g.} by Domain \textit{et
al.} \cite{domain2004kinetic}, but a good parameter set to simulate irradiation
processes in Fe-C under any condition still needs to be elaborated.

Relatively few thoroughly reported and well-documented irradiation experiments
on Fe-C systems can be found in the literature that are actually usable as reference for their simulation with
the OKMC method \cite{malerba2011review}. We
have chosen here two reference experiments: One irradiation
experiment by Eldrup, Singh and Zinkle
\cite{zinkle2006microstructure,eldrup2002dose}
and one post-irradiation annihilation experiment by Eyre \textit{et al.}
\cite{eyre1965electron}.
The first of these
experiments was chosen because of its completeness. The second one is in fact
the only relatively well-documented post-irradiation annealing experiment on Fe
that could be found. Low dose and relatively low irradiation temperature
conditions have been chosen, because they involve less complexity when modelled
with OKMC tools.

Eldrup \textit{et al.} conducted an irradiation experiment on $\alpha$-Fe with about 80 atomic parts per million (appm) carbon and nitrogen at the High Flux Isotope Reactor (HFIR) at ORNL \cite{eldrup2002dose}. They estimate the neutron fluxes to be $4\cdot10^{18}$ n/m$^2$/s ($E > 1$ MeV). The irradiation temperature was around 343 K. From the experiment, they report vacancy cluster number densities, obtained by positron annihilation spectroscopy (PAS) for different
displacements per atoms (dpa) values, up to 0.23 dpa \cite{eldrup2002dose}. They
also report the corresponding number densities of visible SIA clusters from
transmission
electron microscopy (TEM) studies of the same materials
\cite{zinkle2006microstructure}. 

Eyre \textit{et al.} irradiated specimens measuring 76.2 $\times$ 6.35 mm at 333 K for three months up to 0.73 dpa in the DIDO reactor. The thermal and fission neutron doses were $8.2\cdot10^{20}$ and $2.5\cdot10^{20}$ n cm$^{-2}$, respectively. The samples were afterwards annealed for 1 hour at various temperatures up to 773 K \cite{eyre1965electron}. 
Using TEM, they measured the number densities of visible SIA
clusters and their mean diameters at different temperatures. 

To get a more complete picture from the available experimental data in $\alpha$-Fe, we
have also considered other experiments, with less complete data:
\cite{singh1999effects,bryner1966study,robertson1982low,horton1982tem,
takeyama1981}.
A more thorough overview of the available irradiation and annealing experiments
with pure Fe will be published in a separate paper
\cite{malerba2011review}.

In this work, we develop an OKMC model that succeeds in simulating the defect
evolution under irradiation and
post-irradiation annealing of $\alpha$-Fe containing a certain amount of C. Key
in our model is the identification of carbon-vacancy complexes as traps for SIA
clusters, leading to a convincing understanding of the role of carbon on the
nanostructural evolution of iron under irradiation. We will demonstrate
the accuracy of our model by reproducing vacancy and SIA cluster density data
from the above mentioned irradiation and annealing experiments. The paper is
organized as follows: The OKMC
method is described in section \ref{sec:methods}. Sec. \ref{sec:parameterization}
describes our set of parameters for SIA and vacancy clusters in Fe-C systems. In Sec. \ref{sec:FIA}, we show
simulation results under idealized conditions, for the purpose of parameterizing
the traps that are used in the main model. The parameters for the traps are described in Sec. \ref{sec:trap_parameters}. In Sec. \ref{sec:irradiation} and \ref{sec:annealing} are results from
simulations of the nanostructure evolution under irradiation and annealing,
respectively, using our main set of parameters. Finally, we discuss the results and
present our conclusions.

\section{Computation methods}\label{sec:methods}

\begin{figure}
 \centering
  \includegraphics[width=\columnwidth]{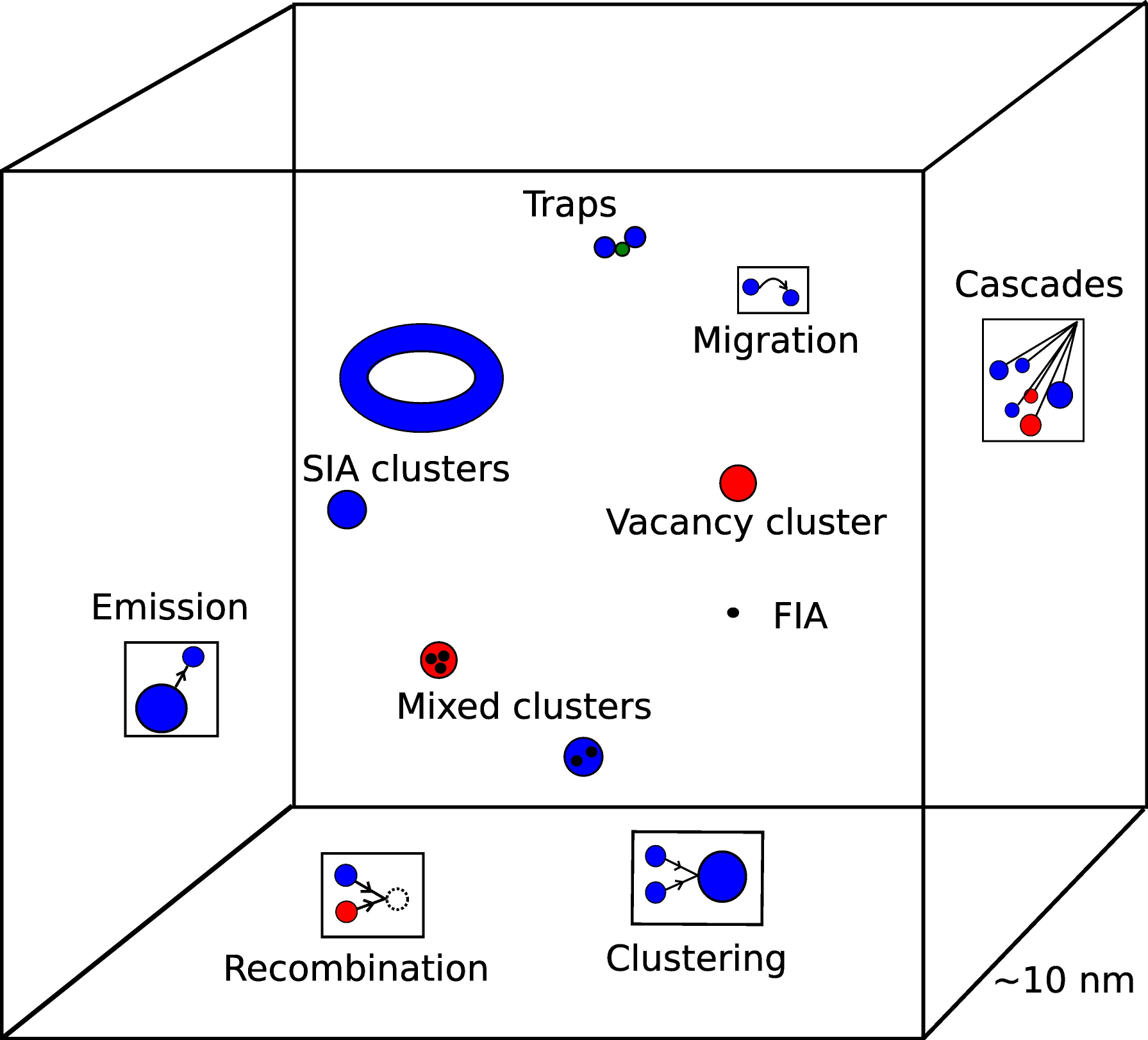}
\caption{The OKMC simulation box with all objects and events used in our model.}
\label{OKMC_simulation_box.pdf}
\end{figure}

For our OKMC simulations, we use the code {\sc LAKIMOCA}, which is already
thoroughly described in \cite{domain2004simulation}. In short, objects and
possible events are introduced in
a simulation box according to pre-defined probabilities (\textit{Cf.} Fig.~\ref{OKMC_simulation_box.pdf}).
Objects may represent vacancy and SIA clusters or any other nano- or
microstructural feature. Every object has an associated reaction volume,
that is generally spherically shaped (except for large dislocation loops, which
have a toroidal shape). When the reaction volumes of two objects overlap, a
predefined
reaction, like clustering between two vacancy clusters or annihilation between a
vacancy and an SIA, takes place.

The events in an OKMC simulation determine the dynamics of the system. The
simplest
event is one object migrating by one step (jump to first nearest neighbour
distance), which is an example of an internal
event. An external event is \textit{e.g.} a cascade, which introduces new vacancy and
SIA objects into the system. Common for all events is that they have a
probability that needs to be defined in the parameterization. In most cases
these probabilities are expressed in terms of frequencies of thermally activated
processes,
\begin{equation}\label{arrhenius}
 \Gamma_i = \nu_i \exp \left(\frac{-A_i}{k_BT}\right),
\end{equation}
where $\nu_i$ is the attempt frequency, alias the prefactor, for event  $i$,
$A_i$ is the corresponding activation energy, $k_B$ is Boltzmann's constant
and
$T$ the temperature. 
For every simulation step, an event is chosen, based on the corresponding
probabilities in the parameterization, using the Monte Carlo algorithm
\cite{metropolis1953equation}. When an event is selected, the progress in
simulation time is calculated using the residence time algorithm
\cite{young1966monte}.
% ,
% \begin{equation}
% \Delta t = \frac{1}{\sum_{i=1}^{N_{int}} \Gamma_i + \sum_{j=1}^{N_{ext}} P_j},
% \end{equation}
% where $P_j$ are the probabilities of the external events, and $N_{int}$ and
% $N_{ext}$ are the total number of internal and external events, respectively.

A second kind of objects employed in our model are traps and sinks. Both
have a spherical reaction volume with a radius that has to be specified. Both may
act on vacancy clusters or SIA clusters, depending on their specification. When
objects interact with a trap, they are bound to the trap by a certain trapping
energy, $E_t$, specified for each kind of trap. 
We use traps to simulate the effect of carbon or carbon-vacancy complexes that
are able to trap SIA clusters. When an object interacts with a sink, it is
absorbed. Sinks are used to allow for the presence of dislocations in the
material. Spherical sinks are used, the radius of which is defined in such a way
that their sink strength equals the sink strength of dislocations. In addition,
the presence of grain boundaries is allowed for, using the algorithm described
in \cite{malerba2007object}. 

% FIA
The foreign interstitial atoms, FIA, are another class of objects. FIA
are a more explicit way to represent carbon, but also 
a more complex one, as more parameters have to be specified.
Indeed, FIA may form mixed clusters with vacancy and SIA. These mixed
objects need to have their own set of parameters as well.

LAKIMOCA can simulate damage production from electron, ion and neutron
irradiation.
In the case of electrons, Frankel pairs are introduced in the
simulation box according to a certain dose rate, following the assumption that
every electron produces one Frenkel pair. The pairs are introduced as
uncorrelated pairs of single vacancy and SIA objects. 

When simulating neutron (or ion) irradiation, debris of vacancy and SIA objects of
different sizes are randomly introduced into the system at a certain rate per
time and volume. The defect populations are chosen randomly from a database
\cite{stoller1996point,stoller1997primary,stoller2000statistical,
stoller2000evaluation,stoller2004secondary,stoller2000role} of displacement
cascades simulated using molecular dynamics (MD) with the Finnis-Sinclair potential
\cite{finnis1984simple}. The database contains cascades with energies of
5 keV, 10 keV, 20 keV, 30 keV, 40 keV, 50 keV and 100 keV. The accumulated dpa is calculated using the NRT formula \cite{domain2004kinetic,norgett1975proposed}:
\begin{equation}
 dpa = \frac{0.8 E_{MD}}{2E_D},
\end{equation}
where $E_{MD}$ is the damage energy, \textit{i.e.} the fraction of the kinetic energy of the primary knock-on atom (PKA) spectrum that is not absorbed by electronic excitation and is well approximated by the energy of the cascades in the MD simulations. $E_D = 40$ eV is the displacement threshold energy for Fe. 

We prefer to use a non-cubic simulation box in order to avoid anomalies from 1D-migrating defects
entering a migration trajectory loop, due to the periodic boundary conditions, as discussed
in \cite{malerba2007object}.

\section{Parameterization for SIA and vacancy clusters}\label{sec:parameterization}

The parameters needed to define the properties of point-defects and their
clusters in our model are listed in Table
\ref{parameters_overview}. All parameters are in principle a function of the
size of the defect clusters, $N^\delta$ (where $\delta=v$ denotes vacancies,
$\delta=i$ SIA and $\delta = f$ FIA). To build these tables, we have gathered
available data from \textit{ab initio}, molecular
dynamics, rate theory, atomistic and OKMC studies and carefully chosen the most
accurate values to be used for our model. The chosen values for all parameters
are described below for all species of objects, vacancies, SIA and traps. 

The number of parameters needed is very large. However, it is important to emphasize that the goal of this work is to establish all physical parameters for which clear indications as to whether their values exist from fundamental studies or experiments, thereby limiting the need to calibrate only a few, if any, parameters, with clear physical interpretation. These, in turn, will be considered satisfactory if their value is physically acceptable.
\begin{table}
\centering
\caption{Overview of all parameters and their annotations: $\delta = v$ for
vacancies, $\delta = i$ for SIA, $\delta=f$ for FIA, $\delta=fv$ for mixed
FIA-vacancy clusters and $\delta=fi$ for mixed FIA-SIA clusters.}
\label{parameters_overview}
\begin{tabular*}{\columnwidth}{@{\extracolsep{\fill}} l c p{0.7\columnwidth}}
 \toprule
$N^\delta$	&  -	& The number of defects in the cluster --- all
parameters that follows are in principle functions of $N^\delta$.\\
\addlinespace
$\nu^\delta$  	& [s$^{-1}$]& The prefactor (or attempt frequency) for the
cluster migration.\\
\addlinespace
$M^\delta$	& [eV]	& The migration energy of the cluster.\\
\addlinespace
$\nu^\delta_d$	& [s$^{-1}$]& The prefactor (or attempt frequency) for a
emission of a defect, $d=i,v$ or $f$, from a cluster.\\
\addlinespace
$M^\delta_d$	& [eV]	& The migration energy of a defect, $d$, emitted from a
cluster.\\
\addlinespace
$B^\delta_d$	& [eV]	& The binding energy of a defect $d$ to a cluster\\
\addlinespace
$r^\delta$	& [Å]	& The capture radius around a given spherical object,
representing its strain field; when two spheres overlap the two objects react
with each other\\
\addlinespace
$\chi^\delta$	&  -	& Parameter determining the shape of the object. If 1,
the
geometrical shape of the cluster is a torus, if 0, the shape is a sphere.\\   
\addlinespace
$\eta^\delta$	& [eV]	& The energy of rotation, used to define a pure
probability (not frequency) of rotation of the Burgers vector associated with
the cluster. A vanishing value corresponds  to fully 3D motion, a value
approaching 1 eV or more corresponds to fully 1D motion.\\
$E^\delta_t$	& [eV]	& The energy by which a defect $\delta$ is bound by a
generic trap.\\
\bottomrule
\end{tabular*}
\end{table}

\subsection{Vacancy clusters} \label{vac_par}

In what follows, we describe in some detail how
the numerical values of the parameters for vacancy-type defects were established.

\subsubsection{Diffusivity}

The diffusion coefficient of a migrating species can be written equivalently as
(see \textit{e.g.} \cite{osetsky2001atomistic})
\begin{equation}\label{eq1}
 D(T) = D_0 \exp\left(\frac{-E_a^D}{k_BT}\right) = f_c(T)\frac{\Gamma(T)d^2_j}{2n},
\end{equation}
where $D_0$ is the diffusivity prefactor, $E_a^D$ is the diffusion activation
energy, $f_c$ ($\sim$1 in the case of three-dimensionally randomly migrating
vacancy clusters ) is the correlation factor (which may depend on
the temperature), $d_j$ is the single jump distance, and $n$ is the
dimensionality of the migration (1 for 1D migration, 3 for 3D migration). The
jump frequency, $\Gamma(T)$ is expressed as \cite{osetsky2001atomistic}:
\begin{equation}
 \Gamma(T) = \nu^v \exp\left(\frac{-E_a^v}{k_BT}\right),
\end{equation}
where $\nu^v$ is the so-called attempt frequency. Assuming $E_a^D \sim E_a^v$
(completely true in case of purely random walk; approximately true so long as
correlated jumps remain rare events), one gets
\begin{equation}\label{eq3}
 D_0 \approx f_c \frac{\nu^v d_j^2}{2n} \approx \frac{\nu^v d_j^2}{2n}.
\end{equation}
Therefore, neglecting correlation
effects, the diffusivity is determined once
attempt frequency and migration energy are known (if correlation effects exist,
they must be introduced in the model separately) and these are the parameters
that must be provided to LAKIMOCA in order to define the mobility of each
migrating species.

% In the case of vacancy clusters, diffusivity and stability for size up to
% $N^v=6$
% have been studied using atomistic KMC methods, based on the use of either
% pre--tabulated or artificial--neural--network-predicted migration energy
% barriers
% as functions of the local environment
% \cite{djurabekova2007stability,pascuet2010submitted}). At the moment this is
% the
% only viable method to study the diffusivity of vacancy clusters as the
% time-frame of MD is too short to allow the same type of study.
% This study provided information about migration attempt frequency, migration
% energy and dissociation energy
% \cite{djurabekova2007stability,pascuet2010submitted}. For higher sizes, while
% waiting for the results of similar studies, reasonable values of migration
% energies and prefactors are used, as described in what follows. It is useful
% to
% emphasize that, even though the parameters proposed here to describe the
% diffusivity of vacancy clusters can of course be improved, the values proposed
% here can be considered fairly established, differently from the case of
% self--interstitial clusters. The only reason to change the mobility parameters
% of vacancy--type defect could be to take into account the effect of alloying
% elements or impurities.

\paragraph{Attempt frequency for migration, $\nu^v$}

For the single vacancy, an attempt frequency of the order of the Debye
frequency, $\nu_1 =
6\cdot10^{12}$ s$^{-1}$ is used. The attempt frequency values
given for the vacancy clusters will be
expressed for convenience
in terms of this value, used as a constant, which was fitted to the
self-diffusion coefficient of iron at 300\textdegree C.

\begin{figure}
 \centering
  \includegraphics[width=\columnwidth]{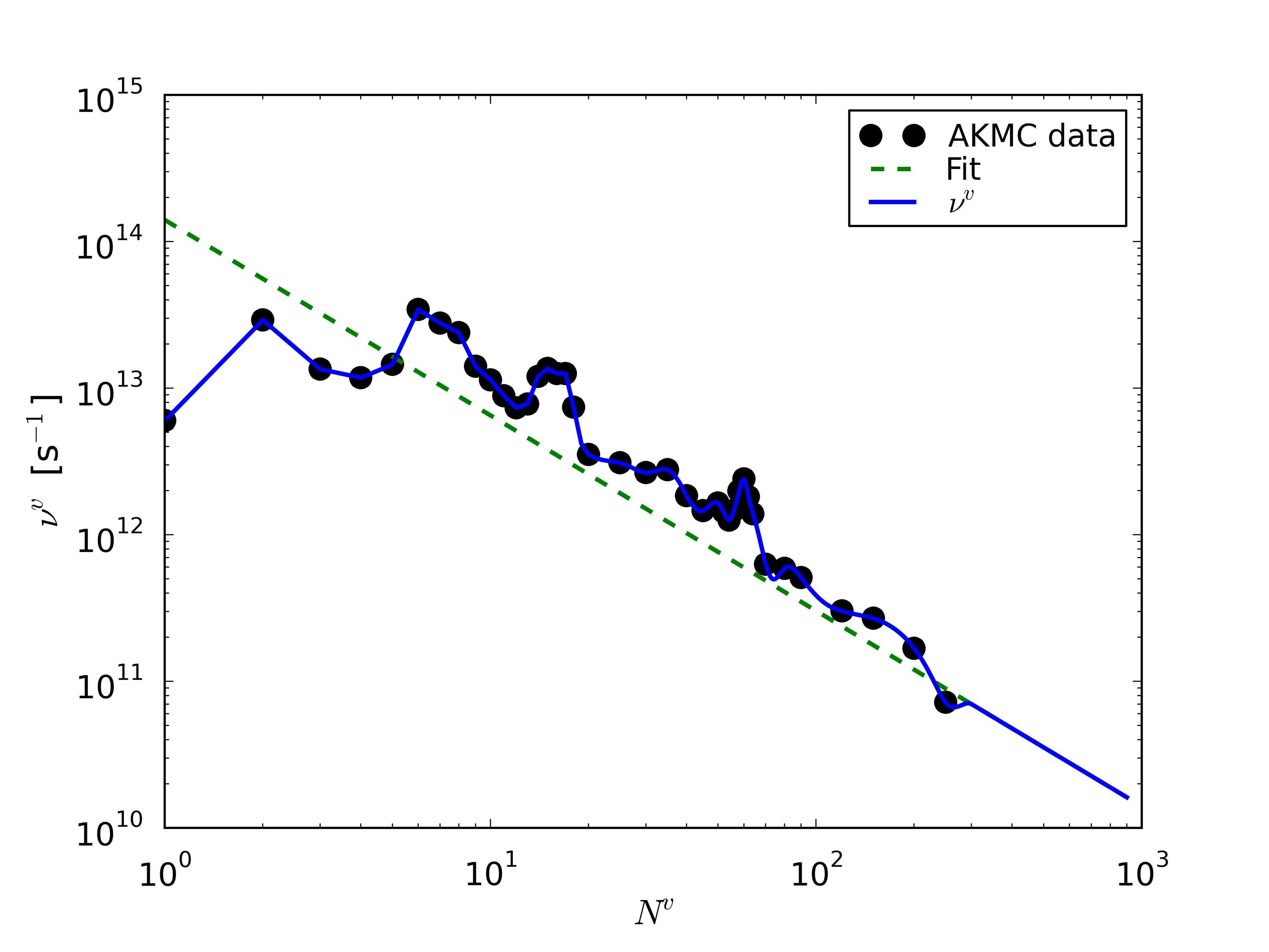}
  \caption{The attempt frequency parameterization, $\nu^v$, interpolated to AKMC
data from \cite{pascuet2011stability,castin2012mobility} and then extrapolated.}
  \label{VAC_nu.pdf}
\end{figure}

For clusters of size $N^v$ = 2--250, values of attempt frequency obtained by
atomistic KMC (AKMC)
\cite{pascuet2011stability,castin2012mobility} are used (\textit{Cf.} Fig.
\ref{VAC_nu.pdf}. Missing values are interpolated using cubic spline. For $N^v >
300$, the clusters are assumed to
migrate by surface diffusion
mechanisms (\cite{nichols1969kinetics}, as in \cite{golubov2007kinetics}):
\begin{equation}\label{eq4}
 \nu^v \sim \frac{\nu_1}{(N^v)^{4/3}}.
\end{equation}
This scaling was smoothly joined to the AKMC values using the correlation:
\begin{equation}
 \nu^v = -1.02064 \cdot 10^{-3} + \frac{1.410626 \cdot 10^{14}}{(N^v)^{4/3}}.
\end{equation}
For clusters of sizes $N^v$ = 251--299, the values are interpolated using cubic
splines.

\paragraph{Migration energy, $M^v$}

The migration energy for the single vacancy is taken to be 0.63 eV, as predicted
by the Mendelev potential \cite{mendelev2003development}, one of the most
reliable for radiation damage
studies in iron. The reason for adopting the inter-atomic potential value instead of the experimental value or the one obtained from density functional theory is
that the relative differences between migration rates
are probably more important than the absolute values. Since the migration
energies for clusters must necessarily be evaluated using the inter-atomic
potential, for consistency, the inter-atomic potential value is also adopted in
the case of the single vacancy. Density functional theory (DFT) calculations for the migration energy of the single vacancy give values between 0.64 eV \cite{becquart2003ab} and 0.67 eV \cite{fu2004multiscale}; experimental values are reported to be 0.55 or 0.57 eV \cite{malerba2010comparison}. Mendelev's potential gives 0.63 eV \cite{mendelev2003development}, which is sort of intermediate between experimental and DFT values. According to DFT, the di-vacancy has a migration energy similar to the single vacancy, whereas the tri-vacancy is more mobile \cite{malerba2010comparison}. This trend is respected by the values from \cite{pascuet2011stability}, that we use. 

\begin{figure}
 \centering
  \includegraphics[width=\columnwidth]{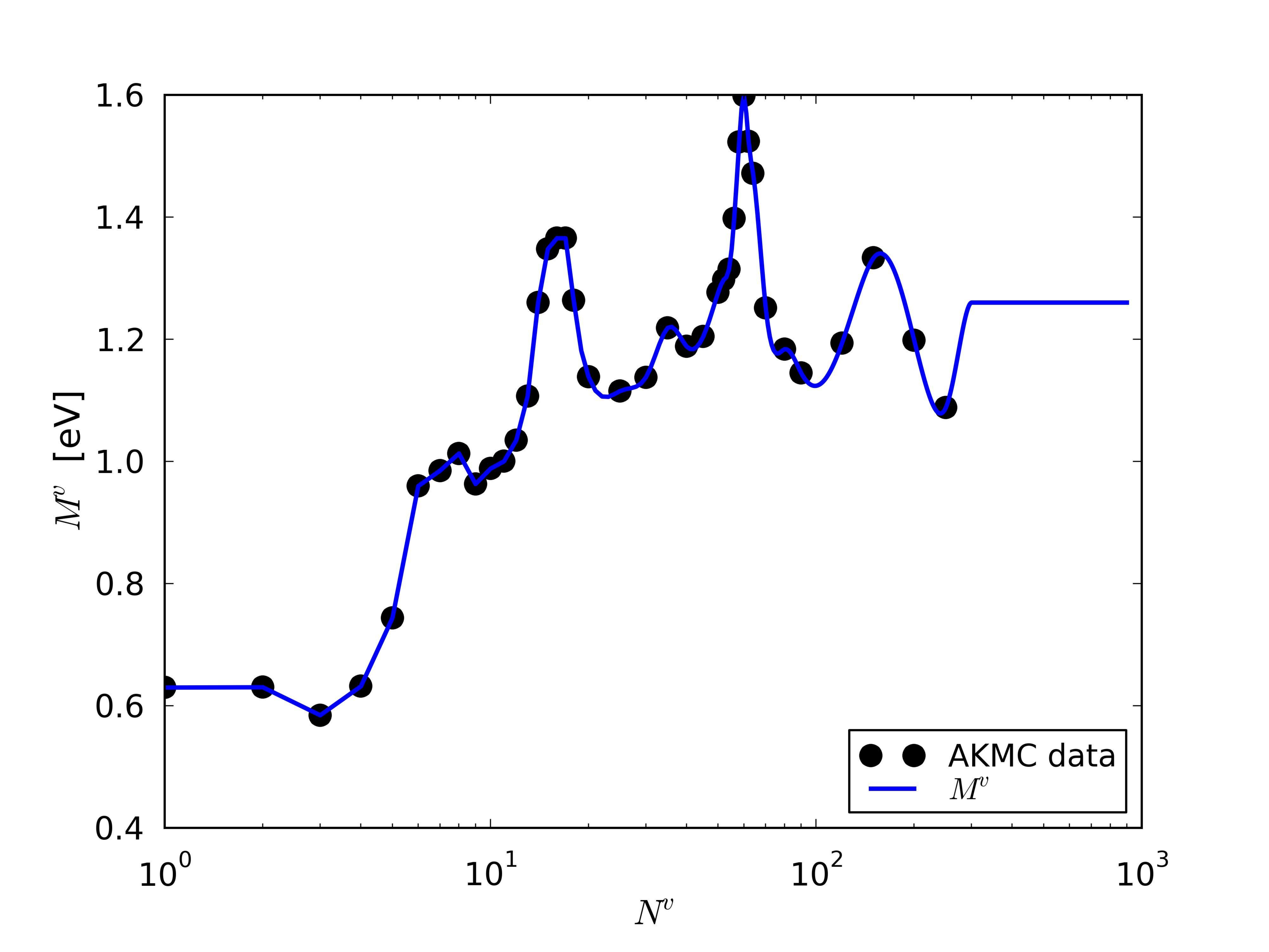}
  \caption{The migration energy for vacancy clusters, $M^v$; interpolated to
AKMC data from \cite{pascuet2011stability,castin2012mobility} and then
extrapolated.}
  \label{VAC_M.pdf}
\end{figure}

For sizes $N^v$ = 2--300, AKMC values \cite{pascuet2011stability} are used and
the missing values are interpolated using cubic splines (\textit{Cf.} Fig.
\ref{VAC_M.pdf}. For $N^v > 300$, a single value is extrapolated from a smooth
fit to the AKMC values, using only cluster sizes $N^v > 10$. Hence, the
obtained migration energy for vacancy clusters of sizes larger than 300
vacancies is set to 1.26 eV. The dataset as a function of $N^v$ is represented
in Fig. \ref{VAC_M.pdf}. Note that the peaks that appear for $N^v < 250$ have a
physical explanation: they correspond to vacancy clusters in which a nearest
neighbour shell is complete: due to their high symmetry, the surface motion of
vacancies in these clusters is more difficult, hence the corresponding migration
energy is higher \cite{castin2012mobility}.

\subsubsection{Dissociation via emission of single vacancies}
Dissociation via emission of clusters is an extremely unlikely event
\cite{castin2012mobility}, so only single vacancy emission is considered. The
dissociation
(or emission) energy is by definition the sum of the migration energy of the
emitted object plus its binding energy to the mother cluster: $E_{diss} =
B^v_v(N^v)+M_v^v(1)$. Here, too, different series of values contribute to
building
the parameterization for all sizes.

\paragraph{Jump frequency for emission, $\nu^v_v$}
In order to be consistent with the rate theory and guarantee that absorption and
emission can reach steady state kinetic equilibrium---not to confuse with
thermodynamic
equilibrium---the vacancy emission must be proportional to the radius of the
cavity, \textit{i.e.} the power $\frac{1}{3}$ of the size expressed in terms of
number of
vacancies, $N^v$:
\begin{equation}
\nu^v_v \sim \nu_1 \sqrt[3]{N^v}
\end{equation}

\begin{figure}
 \centering
  \includegraphics[width=\columnwidth]{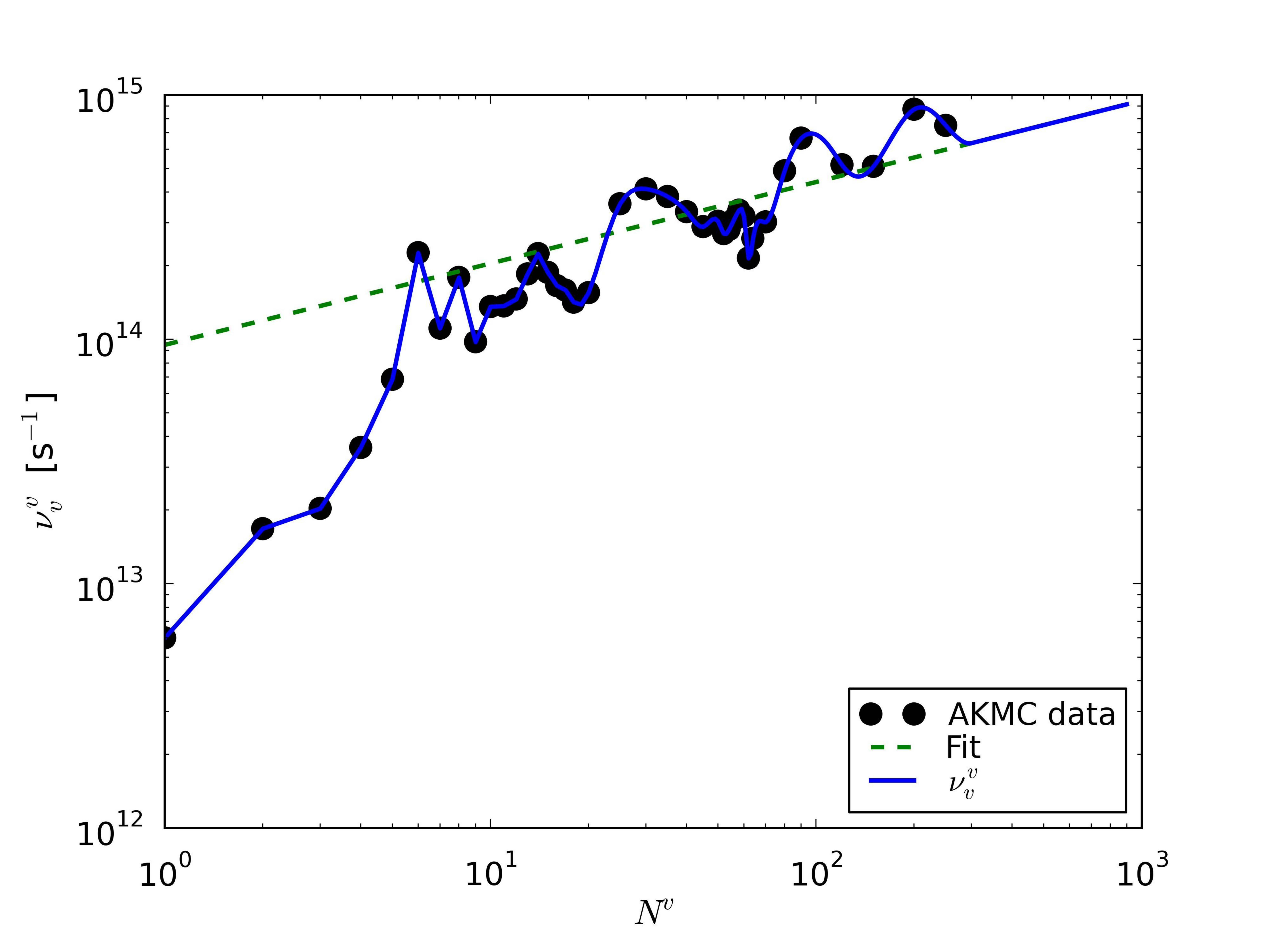}
  \caption{The attempt frequency for emission of a vacancy, $\nu^v_v$, from
clusters of different sizes. The values are interpolated to AKMC data from
\cite{pascuet2011stability,castin2012mobility} and then extrapolated.}
  \label{VAC_nu_e.pdf}
\end{figure}

For $N_v \leq 300$, the parameters were again interpolated to AKMC values
\cite{pascuet2011stability,castin2012mobility} and then extrapolated with the
function (\textit{Cf.} Fig. \ref{VAC_nu_e.pdf}):
\begin{equation}
 \nu^v_v = -6.40388 \cdot 10^{-4} + 9.49849 \cdot 10^{13} \cdot \sqrt[3]{N_v}.
\end{equation}

\paragraph{Migration energy for emission of a single vacancy, $M^v_v$}

This energy is considered independent of the size of the mother cluster, so 
$M^v_v = 0.63$ eV for all $N^v$.

\paragraph{Binding energy, $B^v$}
The binding energy we refer to here is between the emitted single
vacancy and the vacancy cluster.

\begin{figure}
\centering
\includegraphics[width=\columnwidth]{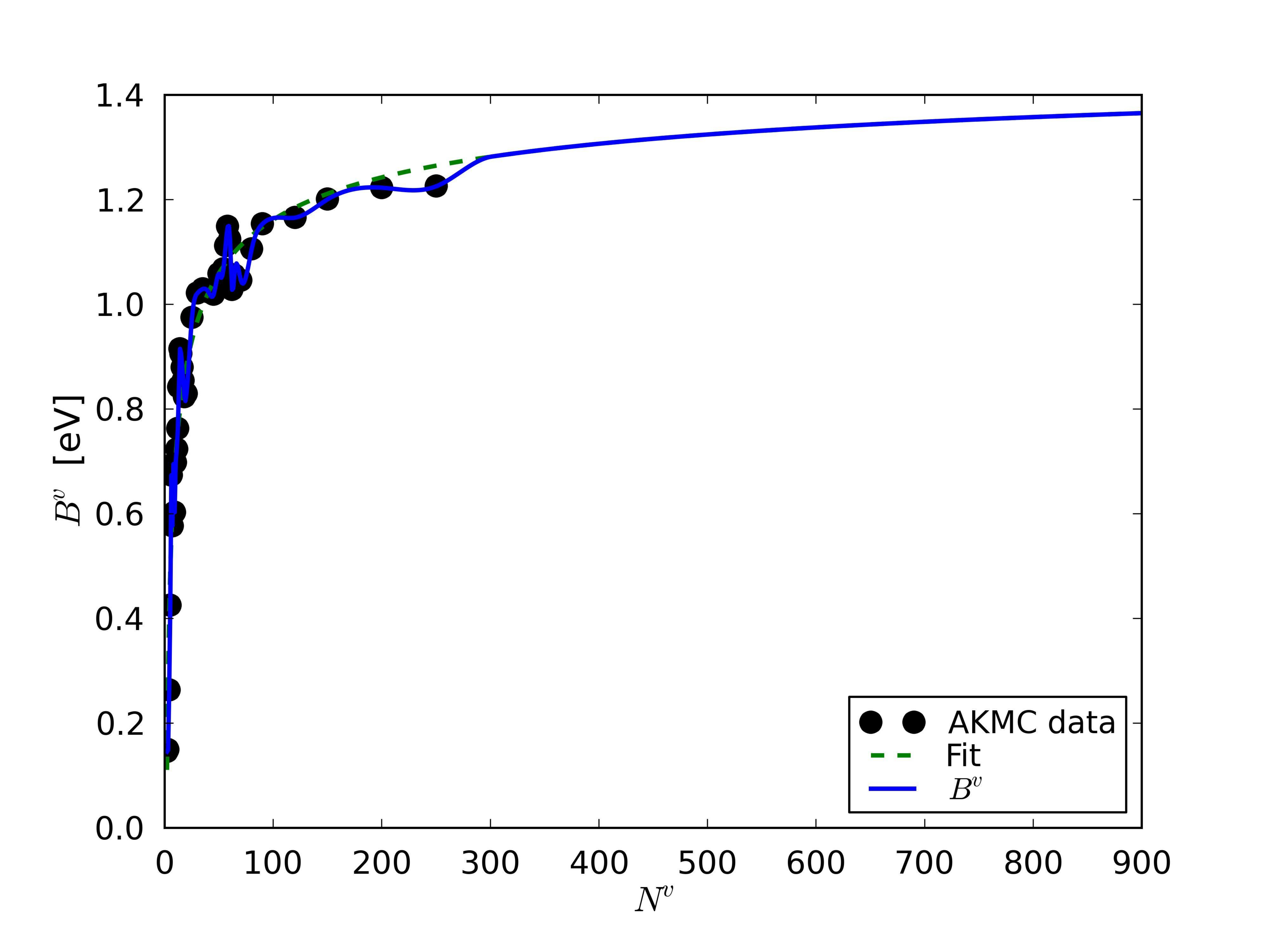}
\caption{The binding energy of a vacancy to a cluster, $B^v$, The values are
derived from dissociation energies, calculated with AKMC
\cite{pascuet2011stability,castin2012mobility} and then extrapolated. The
asymptote of the extrapolated values is the formation energy of a
vacancy according to the interatomic potential used..}
\label{VAC_B.pdf}
\end{figure}

For $N^v \leq 300$. we calculate $B^v = E_{diss} - M^v_v$, from the dissociation
energies
$E_{diss}$, obtained in \cite{pascuet2011stability,castin2012mobility}. These
values are lower than the binding energies statically calculated (either with
the Mendelev potential or by DFT), because these are effective values that
inherently take into account the fact that the configuration of the clusters
keeps changing, thereby making dissociation easier than predicted based on the
energy difference between the dissociated and the fundamental states of the
cluster. In other words, these values include configurational entropy effects on
the dissociation free energy. The missing values were interpolated using cubic
splines. For $N^v > 300$, the values
were extrapolated by fitting the function,
\begin{equation}\label{eq_Bv}
 B^v = 1.71 + 3.39716 [(N^v)^{2/3} - (N^v + 1)^{2/3}],
\end{equation}
to the AKMC values. This equation (\ref{eq_Bv}) interpolates fairly well the
AKMC values up to $N^v = 300$ and extrapolates them especially well, as is shown
in Fig. \ref{VAC_B.pdf}---its asymptote is the formation energy of the vacancy
according to the Mendelev potential, consistently with the fact that the
emission of a vacancy from a very large voids is equivalent to creating a new
vacancy in the bulk by removing an atom from a free surface.

\subsubsection{Other parameters}

\paragraph{The capture radius, $r^v$}

Following Table 2 in \cite{domain2004simulation}:
\begin{equation}
  r^v = \frac{3.3 a_0}{1+ \gamma} + \epsilon + a_0 \left( \frac{3}{8 \pi}
\right)^{1/3} \left((N^v)^{1/3} - 1\right)
\end{equation}
with interstitial bias $\gamma = 1.2$, $\epsilon = 0.01$, and $a_0 = 2.87$ Å (the lattice parameter of iron).
The rationale behind this expression is that the spherical volume associated with
the cluster has a radius given by the radius of the cluster itself (second term,
determined by how many vacancies are contained in it), increased by the range of
the single defect strain field (first term), the latter obtained including the
correction given by removing the radius of the defects on the surface. The value
$3.3a_0$ corresponds to the distance at which recombination between interstitial
and vacancy occurs in iron according to experiments (see refs.~in
\cite{domain2004simulation}). The bias factor $\gamma$ takes into account the
larger strain field of self-interstitials, as compared to vacancies, and appears
here in order to make sure that the recombination distance between SIA and
vacancy is actually $3.3a_0$ (see expression for SIA clusters below: the
interaction distance is the sum of the capture radii of the two objects,
\textit{i.e.} interaction occurs when the two spheres associated with each
object
overlap).

\paragraph{Loop nature, $\chi^v$}

In iron, at least under neutron irradiation, vacancy loop formation is not
expected to be a frequent event, and the most stable clusters are ``spherical'',
so $\chi^v$ = 0 for all $N^v$.

\subsection{SIA clusters}

In what follows we describe in some detail
how the numerical values of the parameters for SIA-type defects were established.

\subsubsection{Diffusivity}

The theory is the same as for vacancies, \textit{i.e.} it is summarised by
Eqs.~\ref{eq1}--\ref{eq3}. Since SIA clusters migrate relatively fast, much
faster than
vacancy clusters, their diffusivity can be studied by MD.
Reliable studies \cite{anento2010atomistic,terentyev2007dimensionality} have
been performed using Mendelev’s potential
\cite{mendelev2003development}, even though it does not reproduce
correctly the stability of the non-parallel configurations of SIA clusters
\cite{terentyev2008self}. Because of the uncertainties
concerning stability and mobility of non-parallel configurations, the actual
parametrization
will not reflect exactly the calculations. Moreover, as in the case of
vacancies, it is justified to change some of the parameters to
explore the importance of specific effects that the MD
simulations cannot take into account (\textit{e.g.} the effect of impurities).
In particular, contrary to the case of vacancies, the migration energy of SIA clusters should not be considered as fully established.

\paragraph{Migration attempt frequency, $\nu^i$, and migration energy,
$M^i$}

For $N^i \leq 7$, the $\nu^i$ and $M^i$ values have been obtained with reference to
Eq.~\ref{eq3}, using as indication values from
\cite{anento2010atomistic}, slightly modified based on a series of
considerations.

Overall there are three categories of migration energy values to be established,
in different ways:
\begin{itemize}
\item Based on experiments \cite{takaki1983resistivity}, the migration energy of
the single SIA in iron is 0.3 eV and the migration energy of the di-SIA is 0.4
eV, fully 3D in both cases. These values are also found in DFT
\cite{fu2004stability}, and the values obtained from dynamic simulations using
Mendelev's potential in \cite{anento2010atomistic}, namely 0.27 and 0.36,
respectively, are also not dissimilar. The di-interstitial is actually more
stable in a
non-parallel configuration \cite{terentyev2008self}; the unfaulting energy,
\textit{i.e.} the energy required to switch from the non-parallel to the
parallel configuration, was however estimated and found to be $\sim$0.4 eV from
the interatomic
potential, suggesting no difference as compared to the migration energy. In a
DFT calculation \cite{polsson} it was found that actually the non-parallel
di-SIA, rather than unfaulting, can migrate remaining in the same configuration,
with an energy of $\sim$0.55 eV. However, here we shall adopt the more
established value of 0.4 eV, which is supported by experiments.

\item The migration energy to be used for clusters with 3--4 SIA should allow
for the fact that these can only migrate when in parallel configurations
(with relatively small energy, \textit{i.e.} 0.14 eV and 0.15 eV, respectively
\cite{anento2010atomistic}) but, according to DFT, will spend most time in
non-parallel configurations; so, the bottleneck for migration becomes the
unfaulting process. The energy involved in the latter is not easy to estimate, because the available
potentials predict the non-parallel configurations to be metastable, so dynamic
simulations are not reliable, while static simulations require guessing all the
steps of a process that, in the case of 3 and 4 SIA, can be fairly complex. It
is, thus, a challenging calculation of uncertain outcome. In the case of the
4-SIA cluster, recently the unfaulting energy seems to have been reliably
estimated from Mendelev's potential to be $\sim$0.8 eV \cite{fan2010unfaulting},
so we shall adopt this as a reasonable value for its effective migration energy
(a previous estimate \cite{
terentyev2008self} had provided 1.7 eV, which seems somewhat too high). In the
case of the 3-SIA clusters the only estimate for the unfaulting energy that is
currently available is $\sim$0.15 eV, which was obtained from Mendelev's
potential in \cite{terentyev2008self}. This value is likely to be an
underestimation, so we tentatively take for the 3-SIA cluster the same effective
value of migration energy as for the di-interstitial, $\sim$0.4 eV.

\item For the migration energy of large clusters ($>6$ SIA), which are assumed
to migrate in 1D mainly, we take the value from MD, $\sim$0.05 eV (See \textit{e.g.} \cite{terentyev2007dimensionality}). Alternatively, different
choices can be made in parametric studies. However, we prefer to attribute the origin of
the higher values found experimentally ($\sim$1.3 eV
\cite{arakawa2007observation}) to the presence of impurities, so here we shall consider fixed the 0.05 eV value and will move to traps the responsibility of effectively changing this value. We have no data for the 5-SIA complex, so we use an intermediate value, 0.1 eV, to have a smoother transition from 0.8 eV for 4-SIA to the value of 0.05 eV for the 6-SIA complex.
\end{itemize}

Taking into account all these considerations, the parameters
tentatively adopted for $N^i \leq 7$ are shown in Table
\ref{SIA_attempt_frequency_table}.
\begin{table}
  \centering
   \caption{The attempt frequencies, $\nu^i$, and the migration energies, $M^i$,
for SIA cluster sizes of $N^i$ = 1--7.}
\label{SIA_attempt_frequency_table}
\begin{tabular*}{\columnwidth}{@{\extracolsep{\fill}} l c c l}
 \toprule
$N^i$	& $\nu^i$& $M^i$	  &  	\\
	& [$10^{13}$ s$^{-1}$]& [eV]&	\\
\midrule
1	& 8.071 & 0.31 & (Exp.~value)	\\
2     	& 34.15 & 0.42 & (DFT and exp. value)	\\
3     	& 1.175 & 0.42 &		\\
4     	& 1.195 & 0.8  &		\\
5     	& 0.156 & 0.1  &		\\
6     	& 0.156 & 0.05 &		\\
7	& 0.171	& 0.05 &		\\
\bottomrule
\end{tabular*}
\end{table}
Concerning the prefactors for $N^i \geq 7$, the following expression is used:
\begin{equation}\label{eq29}
 \nu^i = \frac{c}{(N^i)^{0.8}},
\end{equation}
Here, $c$ is determined by requiring Eq.~\ref{eq29} to give the same values for
$N^i=7$ as the values in the table above. Thus, $c=8.11\cdot 10^{12}$. The
exponent of $N^i$ in the denominator can theoretically vary between 0.5
(independent crowdion model) \cite{barashev2001reaction} and 1 (migration via
kink pair formation along the edge of a loop) \cite{wirth2000dislocation}. The
value 0.8 adopted here respects the theory and corresponds to the value
determined
experimentally by Arakawa \textit{et al.} \cite{arakawa2007observation}.

\subsubsection{Dissociation via emission of single SIA}

The dissociation (or emission) energy, $E_{diss}$, is defined as the sum of
the migration
energy of the emitted object plus its binding energy to the mother cluster:
$E_{diss}=B^i_i(N^i)+M^i_i(1)$. SIA clusters are known to be thermally highly
stable, \textit{i.e.} dissociation by emission of single SIA is a
very unlikely event, due to the strong binding energy. Therefore, \textit{a
priori}, any
large enough value is acceptable and only small clusters (size 2--3) will
actually have a small probability of emitting. For this reason, it is not
really worth devoting specific studies to refine parameters in this case.
However, recently binding energy values calculated using Mendelev’s potential
have become available and could therefore be used \cite{abe2009lowest}. Here we
take a simplified approach, as follows.

\paragraph{Jump frequency for emission, $\nu^i_i$}

The jump frequency for emission is arbitrarily assumed to be the
same as the one for the migration of the single vacancy and independent of the
mother cluster size: 
\begin{equation}
\nu^i_i = \nu_1 = 6\cdot10^{12} \, \textrm{s}^{-1}
\end{equation}
for all $N^i$.

% In principle other choices could be made, \textit{e.g.} using the attempt
% frequency for
% the
% migration of the single interstitial or introducing a sensible dependence on
% the
% mother cluster size, but due to the unlikeness of the event, these different
% choices would have very little impact on the results.

\paragraph{Migration energy for emission of a single SIA, $M^i_i$}

The migration energy for one SIA is simply chosen to be equal to the migration
energy of the single interstitial, \textit{i.e.} $M^i_i = M^i = 0.3$ eV for all
$N^i$.

\paragraph{Binding energy, $B^i_i$}

The binding energy we refer to here is the one between the emitted single SIA
and the SIA cluster. The simplest option is to use the formula reported
\textit{e.g.} in
\cite{domain2004simulation}. For $N^i>1$:
\begin{equation}
 B^i_i(N^i) = e_{for} + \frac{(B^i_i(2)-e_{for}) ( (N^i)^s -
(N^i-1)^s)}{2^s-1},
\end{equation}
where $s=\frac{2}{3}$, $B^i_i(2) = 1.0$ eV (binding energy of the
di--interstitial);
and $e_{for} = 4.0$ eV (formation of the single interstitial).

\subsubsection{Other parameters}

\paragraph{The capture radius, $r^i$, and the loop nature, $\chi^i$}

The capture radius for an SIA clusters is determined having in mind the fact
that the SIA in the cluster are parallel to each other and distributed on a
(111) plane, forming a platelet that can be approximated by a round disc for large enough
sizes. The sphere associated with the cluster is assumed to be the one whose
equatorial plane is the habit plane of the cluster, \textit{i.e.} the disc
associated, and whose equator is the edge of the loop.
The capture radii are thus given by
\begin{eqnarray}\label{eq_ri}
 r^i &=& r^i_0+\frac{a_0}{\sqrt{\pi \sqrt{3}}}(\sqrt{N^i}-1),\\
 r^i_0 &=& \gamma \frac{3.3a_0}{1+\gamma}\label{eq_ri0}
\end{eqnarray}
where $\gamma=1.2$ (interstitial bias) and $a_0 = 2.87$ Å is the lattice
parameter of iron. 

For large SIA clusters, we know that the strain field of a perfect dislocation loop is limited to a region surrounding the edge dislocation delimiting it, while a perfect lattice with no strain field is found
at the centre of it. The shape of the strain fields of larger loops are thus better
approximated by a torus. Tentatively, we assume that SIA clusters become loops when they exceed size 150.
So, for $1 < N^i < 150$, the objects are spherical, denoted by $\chi^i$ = 0,
and for $N^i \geq 150$, they have toroidal form, denoted by $\chi^i$ = 1. The major toroidal radius, $R$, is then given by 
\begin{equation}
R = \sqrt{\frac{a_0^2 N^i}{\pi\sqrt{3}}}.
\end{equation}
This expression comes from the assumption that the toroid is approximated by
a circle with an area equal to $N^i$ atomic areas. The minor toroidal radius, $r=7.215$
Å, is given by Eq. \ref{eq_ri} for $N^i=2$ for all sizes of
toroidal SIA clusters.

\paragraph{Dimensionality of migration, $\eta^i$}

Small clusters may flip their Burgers vector as a consequence of disturbances
produced by impurities or simply as a spontaneous thermally activated process.
The existence of non-parallel configurations and the possibility that, while
migrating, the cluster turns into one of them can be an additional mechanism,
because when unfaulting, the Burgers vector may be different from the previous
one. This effect is taken into account in the code by including a rotation
energy from which a pure probability (not a frequency) is derived, in terms of a
Boltzmann expression. A rotation energy $\eta^i=0$ produces changes
of
the Burgers vector (= jump direction) at each jump, thereby corresponding to
fully
3D migration. $\eta^i > 0$ reduces the chances of changing direction at each
jump,
so that the defect changes jump direction only after a number of 1D jumps.
Values of $\eta^i$ in excess of 1 eV provide almost fully 1D migration
(depending on
the temperature).

Precise information about the values of $\eta^i$ and the actual dimensionality
of the migration of the different clusters is not available, but it is known
that there is a transition from fully 3D to fully 1D migration with increasing
size. We use $\eta^i = 0$ for single and di-interstitials and increase the value
gradually to 1 eV for larger sizes.
% 
% 
% 
% \begin{verbatim}
% N       E_rot [eV]
% ------------------
% 1       0
% 2       0
% 3       0.1
% 4       0.2
% 5       0.3
% 6       0.4
% 7       0.5
% 8       0.6
% 9       0.7
% 10      0.8
% 11      0.9
% 12      1.0
% \end{verbatim}
% For $N^i > 12$, $\eta^i=1.0$ eV.

\subsection{Treatment of grain boundaries and dislocations} \label{grains_disl}

Grain boundaries and dislocations are sinks for both SIA and vacancies. In
LAKIMOCA, a defect is removed when it has travelled farther than the length of
the average grain size \cite{domain2004simulation}. 

Dislocations are here assumed to be sinks only for small defects. As the
dislocation density in the material by Eldrup \textit{et al.}
\cite{eldrup2002dose} is very low, $\rho_d = 10^{12}$ m$^{-2}$, by applying the equivalence with spherical sinks \cite{nichols1978estimation}, \textit{i.e.}
\begin{equation}\label{spherical_ss}
 k^2_s = 4\pi r_s n_s = \rho_d,
\end{equation}
we found that $r_s$, the radius of the sphere, has a very small value, assuming that the number density of spherical sinks, $n_s$ to be 1 sink in the system. As a consequence, the influence of the dislocation bias on the results of the model cannot be assessed when comparing with this reference experiment.

\section{Study of populations of small C-V complexes} \label{sec:FIA}
  
In this section we report the results of a study of post-irradiation annealing
performed under ideal conditions. Namely, we consider only vacancies and FIA (C atoms). The purpose of this
study is to have an idea of which C-V complexes should be expected to form in
majority after irradiation at $<$370 K and how the population of complexes would
change upon annealing up to above 600 K. This study is then used as guideline for the parameterization of traps, described in section \ref{sec:trap_parameters}. For the sake of simplicity, we
restricted the degrees of freedom of the system by only allowing V$_1$, V$_2$, C, CV, CV$_2$ and
C$_2$V complexes to be stable, as it makes it possible to compile a complete parameter set
based on available data, without any need for fitting. Moreover, indications from
interatomic potentials seem to suggest that larger complexes might not be
especially stable, although uncertainties remain in this respect 
\cite{terentyev2011interaction}. The parameters used for this specific simulation 
are described in the next section.

\subsection{Parameterization with explicit C objects}

For vacancies, the same parameters were used as described before in Sec.
\ref{vac_par}, except that the migration energy for emission is set to $M^v_v =
0.0$ for
vacancy objects with more than two vacancies ($N^v>2$), in order not to form voids and focus the attention on the relative fraction of small C-V complexes, which are expected to be the most common.

For single carbon, the attempt frequency is set to the standard value
$\nu^f=\nu^1=6\cdot10^{12}$ s$^{-1}$ and the migration energy to $M^f = 0.90$
eV \cite{domain2004ab}. The capture radius is set to 5.0 Å. Complexes of only
carbon
are forbidden by
making them immobile, with a very high migration energy $M^f \geq 10.0$ eV, and
unstable,
with a binding energy of a carbon atom to the cluster, $B^f_f = 0.0$, for
complexes with more than one carbon, $N^f > 1$. The parameters for mixed C-V
clusters are shown in Table \ref{FV_parameters_table}.
\begin{table}
  \centering
   \caption{The parameters for the FIA and vacancy-FIA objects for the study of
populations of small C-V complexes. Only V$_1$, V$_2$, C, CV, CV$_2$ and
C$_2$V are allowed. The parameters make all other complexes unstable by
putting the binding energy $B^{fv}_v=0$. $\nu^{fv} = \nu^{fv}_{f} = \nu^{fv}_{v}
= 6.0\cdot10^{12}$ s$^{-1}$, $M^{fv} = \infty$, $M^{fv}_{f} = 0.902$ eV
\cite{domain2004ab} and
$M^{fv}_{v} = 0.63$ eV for all $N^f$ and $N^v$.}
\label{FV_parameters_table}
\begin{tabular*}{\columnwidth}{@{\extracolsep{\fill}} l l l l l}
\toprule
Complex 	& $N^f$	&  $N^v$& $B^{fv}_f$	& $B^{fv}_v$	\\
		&	&  	& [eV]		& [eV]		\\
\midrule
CV 		& 1 & 1 & 0.64 \cite{forst2006point}  & 0.64 \cite{forst2006point}	\\
CV$_2$ 		& 1 & 2 & 1.01 \cite{becquart2011p60} & 0.49 \cite{forst2006point}	\\
CV$_3$  	& 1 & 3 & 0.93 \cite{becquart2011p60} & 0.0	\\
CV$_4$  	& 1 & 4 & 0.96 \cite{becquart2011p60} & 0.0	\\
CV$_5$  	& 1 & 5 & 0.23 \cite{becquart2011p60} & 0.0	\\
CV$_6$  	& 1 & 6 & 1.20 \cite{becquart2011p60} & 0.0	\\
CV$_n$  	& 1 & $7\leq$ & 1.20 & 0.0	\\
\midrule
C$_2$V  	& 2 & 1 & 1.01 \cite{forst2006point} & 2.3 \cite{forst2006point}	\\
C$_2$V$_2$  	& 2 & 2 & 1.18 \cite{forst2006point} & 0.0	\\
C$_2$V$_3$  	& 2 & 3 & 0.93 & 0.0	\\
C$_2$V$_4$  	& 2 & 4 & 1.01 & 0.0	\\
C$_2$V$_5$  	& 2 & 5 & 1.23 & 0.0	\\
C$_2$V$_6$  	& 2 & 6 & 1.30 & 0.0	\\
C$_2$V$_n$  	& 2 & $7\leq$ & 1.30 & 0.0	\\
\midrule
C$_3$V  	& 3 & 1 & 0.00 & 2.3	\\
C$_3$V$_2$  	& 3 & 2 & 0.00 & 0.0	\\
C$_3$V$_n$  	& 3 & $3\leq$ & 0.00 & 0.0	\\
\midrule
C$_4$V  	& 4 & 1 & 0.00 & 2.3	\\
C$_4$V$_2$	& 4 & 2 & 0.00 & 0.0	\\
C$_4$V$_n$	& 4 & $3\leq$ & 0.00 & 0.0	\\
\bottomrule
\end{tabular*}
\end{table}

\subsection{Results} \label{sec:CV_results}

We introduced a random population of carbon in a simulation box of size 
$150\times 200\times 250\times a_0^3$. We included dislocations as sinks in a
similar way to what was described in Sec. \ref{grains_disl}. We also used a
grain size of 33 $\mu$m. 

\begin{figure*}
 \centering
 \begin{subfigure}[b]{\textwidth}
 \centering
  \includegraphics[width=0.5\textwidth]{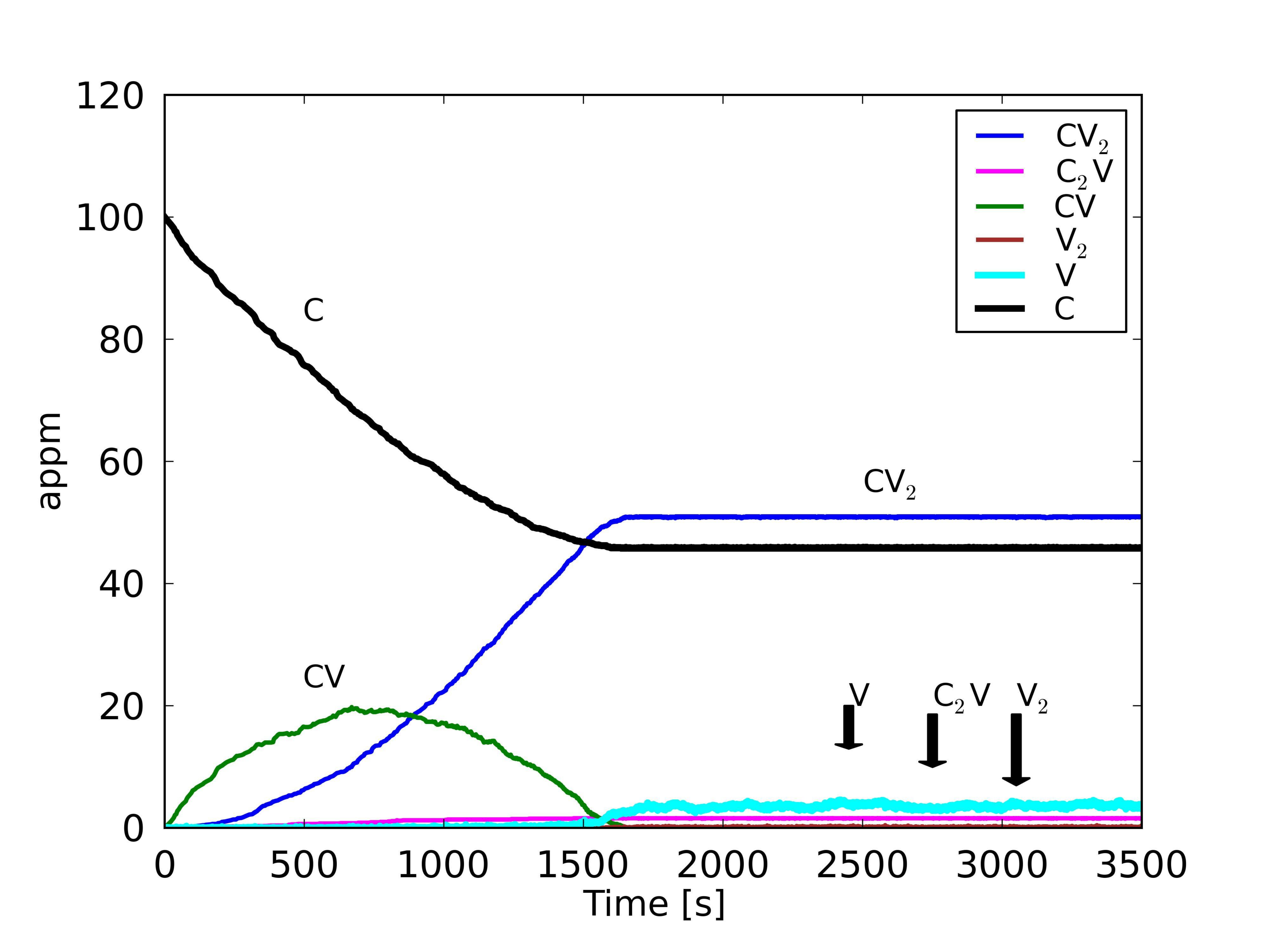}
  \caption{}
  \label{R20120521-00.pdf}
\end{subfigure}
\begin{subfigure}[b]{\textwidth}
\centering
  \includegraphics[width=0.5\textwidth]{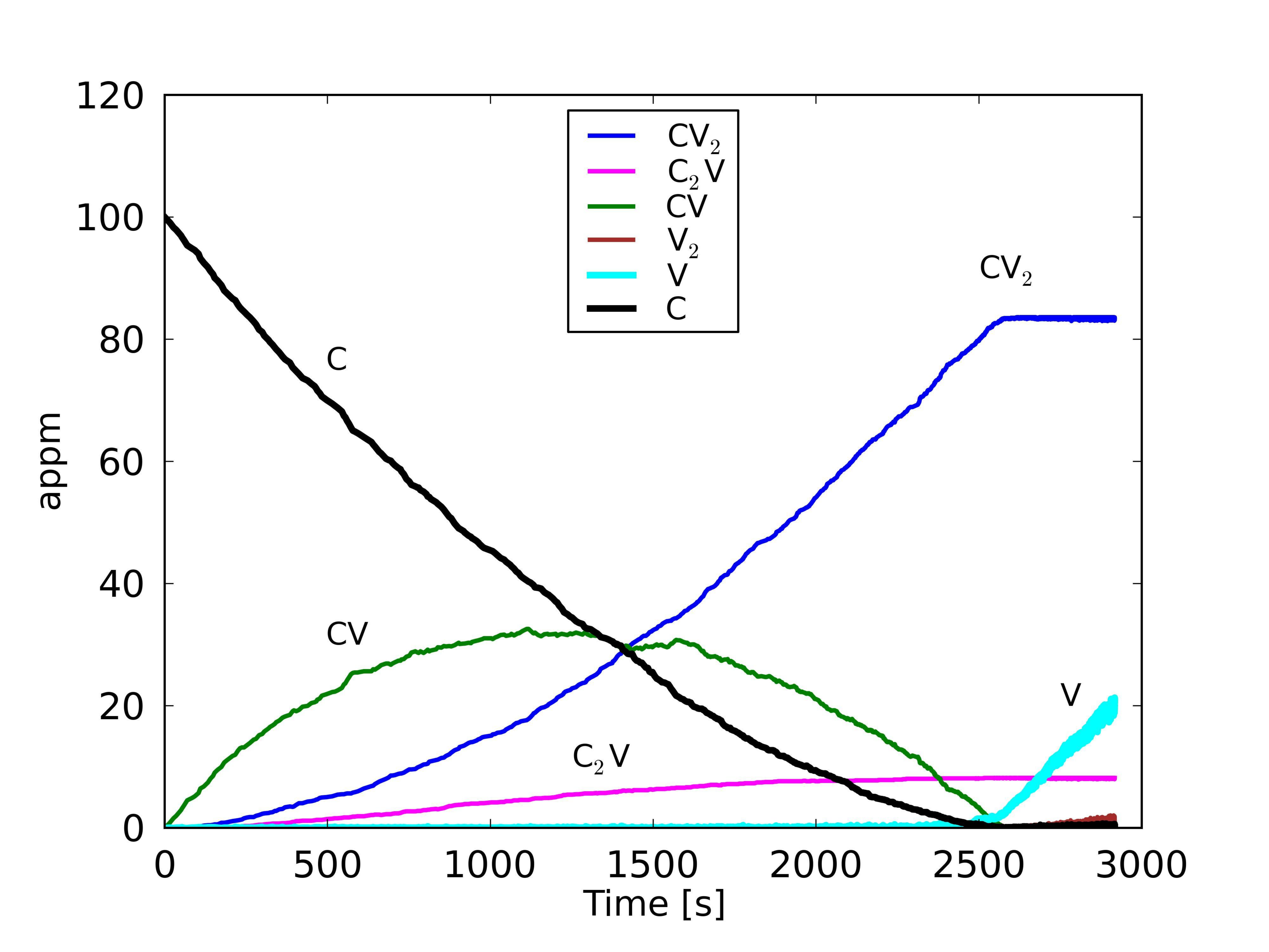}
  \caption{}
  \label{R20130128-02.pdf}
\end{subfigure}
  \caption{The evolution of a carbon-dominated (a) and a vacancy-dominated system (b) with small C-V complexes and under irradiation.}
  \label{R20111013-1.pdf}
\end{figure*}

The irradiation of the system at 333 K was simulated by a constant flux of one
vacancy per second. As the system could possibly behave very differently if it contains more vacancies than C atoms, or if the populations are more equal, we used one case were a number of vacancies were systematically removed continuously (by artificially disappearing at grain boundaries, beyond the correct sink strength) and one case were they were not. The evolution of the density of the different complexes is shown for both regimes in Fig.~\ref{R20111013-1.pdf}. Both systems reach a steady state, regarding the C-V complexes. In the system with low amount of vacancies (in the end an equal amount of C and V remains), the steady state is reached faster and not all C atoms are bounded to vacancies. In the final state, the
number of CV$_2$ complexes is 32 times higher than the number of C$_2$V. The CV complexes have
disappeared completely by becoming C$_2$V. In the case with high amount of V (in the end twice as many V as C), all C are bounded to V and the level of CV$_2$ is about ten times higher than the density of C$_2$V. 

\begin{figure*}
\centering
\begin{subfigure}[b]{\textwidth}
\centering
  \includegraphics[width=0.5\columnwidth]{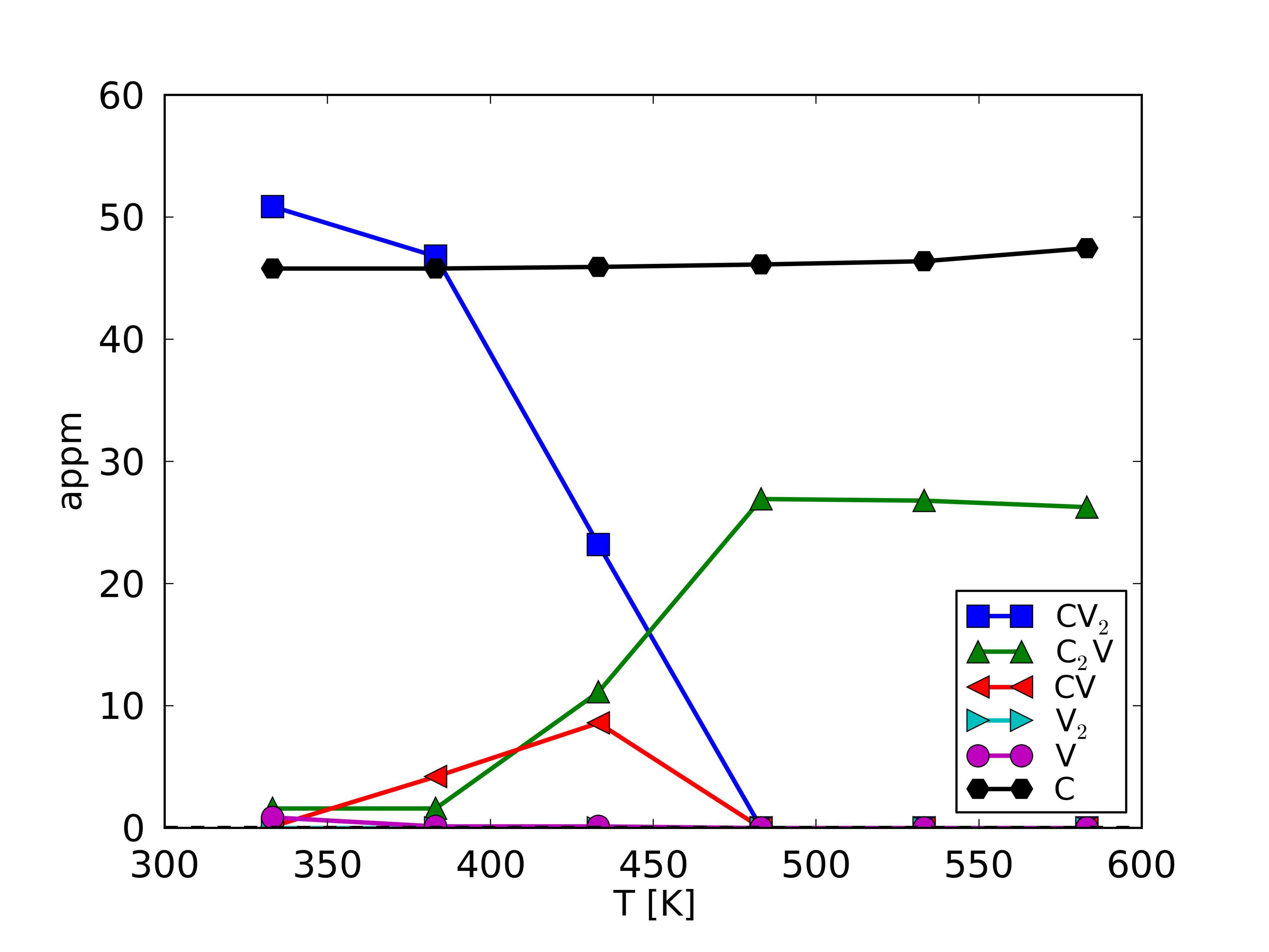}
  \caption{}
\end{subfigure}
\begin{subfigure}[b]{\textwidth}
 \centering
  \includegraphics[width=0.5\columnwidth]{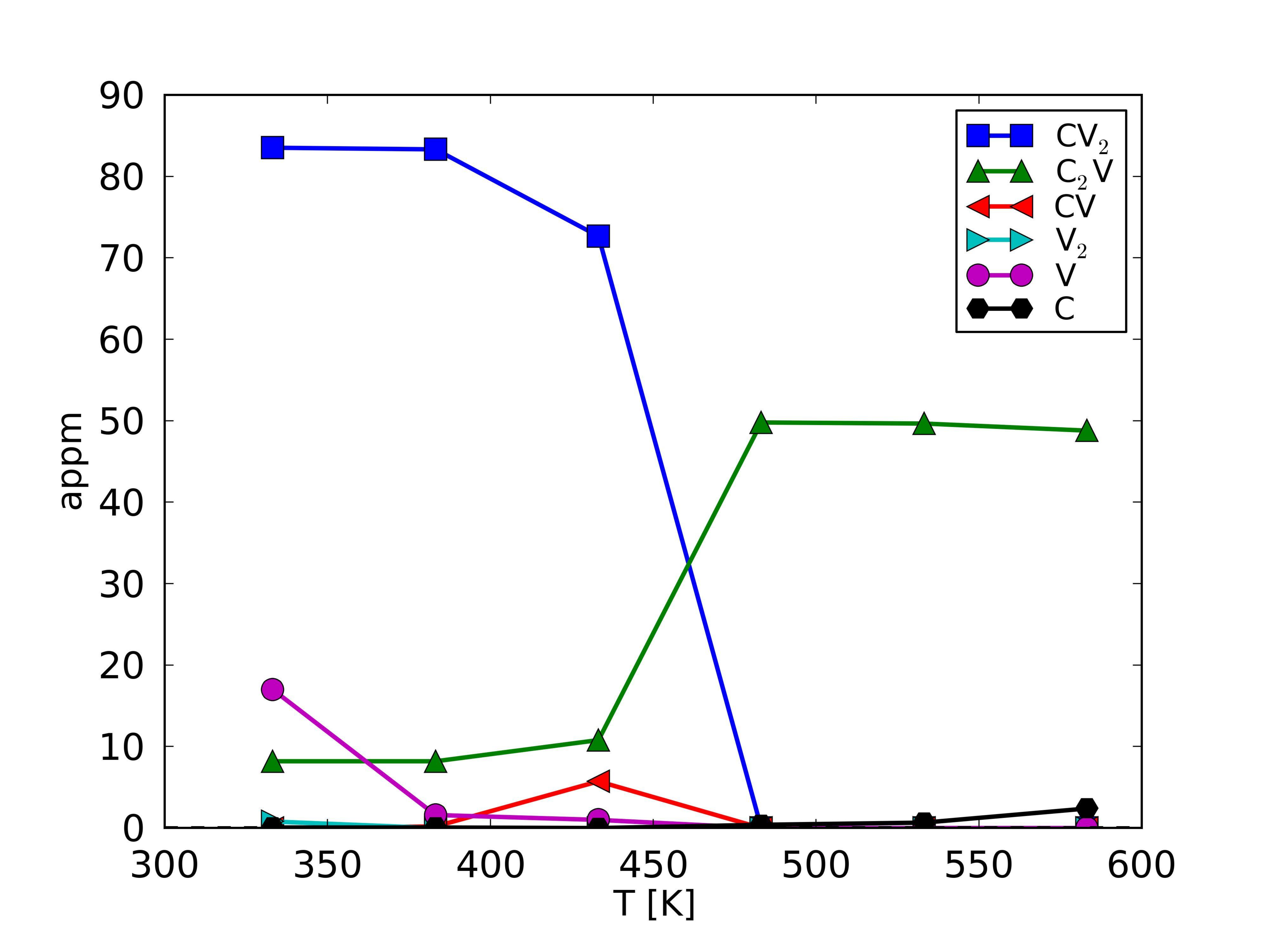}
  \caption{}
  \label{R20130128-20_323484.pdf}
\end{subfigure}
\caption{Annealing of the small complex carbon-dominated (a) and vacancy-dominated (b) system to obtain the temperature
dependence.}
\label{R20111102-1.pdf}
\end{figure*}

The previously irradiated systems were then annealed by raising the temperature
by steps of 50 K, keeping the system at each temperature for one hour. The
evolution of the density of the complexes is presented in
Fig.~\ref{R20111102-1.pdf}. It is observed that CV$_2$ complexes are
dominating at low temperature, below 480 K, whereas the C$_2$V
complexes dominate at higher temperatures for both cases. The major difference between the two regimes is the absence of free C in the system with high amount of V. 

\section{Parameters for the effect of carbon}\label{sec:trap_parameters}

Foreign interstitial atom (FIA) objects are the most explicit way of describing
carbon, but because of the formidable complexity in terms of parameters that
are added by the complexes it forms with point-defect clusters, we used them only
in Sec. \ref{sec:FIA}. In this section, we describe the parameters for the traps than translate the effect of C and which, together with the set of parameters for SIA and vacancies, described in Sec. \ref{sec:parameterization}, will form the full model. The traps represent
not only single C atoms, but also small carbon-vacancy clusters, such as CV, CV$_2$ and C$_2$V. The latter
complexes bind strongly enough to SIA clusters to be able to trap them. The traps are
essential for the simulations to reproduce the experimental data.

CV, CV$_2$ and C$_2$V complexes have been found to be highly stable by MD
simulations 
\cite{terentyev2011interaction} and in particular the C$_2$V complex is stable
up to 700 K. There is also ample experimental evidence of the existence of these
complexes, as reviewed in \cite{malerba2011review}. MD
simulations by N. Anento and A. Serra \cite{anento2013carbon}
show that C-V complexes bind to $1/2\langle111\rangle$ SIA clusters with
different binding energy, depending on whether the complex is interacting with
the edge or the centre of the SIA cluster. At the edge, the vacancies in the complex
tend to recombine, whereas this does not happen at the centre. The binding
energies for C-V complexes with $1/2\langle111\rangle$ SIA clusters according to
MD simulations are given in Table \ref{CV_SIA_binding_E}. It can be seen that
CV$_2$ and C$_2$V are particularly strong traps for SIA clusters, depending on the point of interaction and the temperature. CV$_2$ is a strong trap if it interacts with the centre of the SIA cluster, but weak if interacting with the edge. All complexes with only one C are weak traps if interacting with the edge as the V will recombine. CV$_2$ is a strong trap only if it interacts with the centre.
\begin{table}
\centering
\caption{Binding energies between C-V complexes and the edge or the centre 
of large SIA clusters (size 61 SIA), according to MD simulations \cite{anento2013carbon} .}
\label{CV_SIA_binding_E}
\begin{tabular*}{\columnwidth}{@{\extracolsep{\fill}}l l l}
\toprule
	& \multicolumn{2}{c}{SIA binding energy [eV]}\\
	\cmidrule{2-3}
 	& Centre	& Edge	\\
\midrule
C	& 0.0		& 0.6 \cite{terentyev2011interaction}\\
CV	& 0.3		& 0.75 \\
CV$_2$	& 1.4		& 0.8 \\
C$_2$V	& 0.4		& 1.4--1.5\\
\bottomrule
\end{tabular*}
\end{table}

We use generic spherical traps to simulate not only the formation of C-V
complexes, but also their interaction
with SIA clusters. Gyeong-Geun Lee \textit{et al.} \cite{lee2009kinetic} noted
that it is crucial that the efficiency of the SIA traps depends on the size of
the trapped SIA cluster, as the SIA clusters otherwise tend to merge together until
only a single large SIA cluster is left in the simulation box, independently of
the size of the simulation box. To obtain this dependency, Lee \textit{et al.}
used a dissociation rate that was proportional to $(N^i)^{-1}$ or $(N^i)^{-2}$. Here, we adopted a
method where we specify the trapping energy $E^i_t$ explicitly for all sizes, including temperature dependence when necessary. In Table \ref{SIA_trapping_energies} the trapping energies for SIA and vacancy clusters up to size 6 are presented. 
%For sizes 7--$N_{th}$, $E_t^i = 0.6$ eV. Above the threshold size, $N_{th}$, $E_t^i$ depends on the temperature, such that $E_t^i = 1.2$ eV at 330--480 K, $E_t^i=1.4$ eV at 480--680 K and $E_t^i = 0.6$ K above 680 K. For vacancies, $E_t^v = 0.0$ eV for all sizes above $N^v=6$. We used a capture radius of 5.0 Å for all traps.
\begin{table}
\centering
\caption{The trapping energies for the SIA traps. The SIA values for sizes 1--4
and the vacancy values for sizes 1--6 are from DFT calculations
\cite{becquart2011p60}.}
\label{SIA_trapping_energies}
\begin{tabular*}{\columnwidth}{@{\extracolsep{\fill}} l c c c c}
\toprule
$N^i$ 	& SIA $E^i_t$	& Vac. $E^v_t$\\
	& [eV]		& [eV]\\
	\midrule
1 	& 0.17		& 0.65\\
2 	& 0.28		& 1.01\\
3 	& 0.36		& 0.93\\
4 	& 0.34		& 0.96\\
5	& 0.60		& 1.23\\
6	& 0.60		& 1.20\\
% 7--$N_{th}$	& 0.60		& 0.60		& 0.60		& 0.00\\
% $N_{th}<$& 1.20		& 1.40		& 0.60 		& 0.00\\
\bottomrule
\end{tabular*}
\end{table}

For simulations at low temperature, 330--480 K, the sizes of SIA clusters may
be divided into three
categories, in order to define their trapping energy. For size 1--4, we use DFT
data for the binding energy between a C atom and an SIA cluster, according to \cite{becquart2011p60}. From size 5 to the threshold
size, $N_{th}$, we use the value of the binding energy for a C atom to a SIA
cluster, 0.6 eV \cite{terentyev2011interaction}. This value is the same whether the cluster interacts with a single C atom or a CV$_2$ complex, which is the dominating trap in this range of temperatures. Indeed, for clusters sufficiently small, the highest chance is that the
interaction occurs with the edge, leading to recombination of the vacancy and
interaction energy with single C, \textit{i.e.} 0.6 eV. Above size $N_{th}$, in
the third category, we use a strong binding energy, 1.2 eV, that can be associated to the CV$_2$ complex bound to the centre of the SIA, and comparable to MD results in \cite{anento2013carbon}. The threshold, $N_{th}$, is used as a calibration parameter,
that effectively takes into account the probabilities for three different interactions with a SIA
cluster at 330--480 K: the strong interaction of CV$_2$ complex with the
centre, the weak interaction of ditto complex with the centre or the weak interaction of a
single C with an SIA cluster. 

For temperature of 480--680 K, we assume that the dominating strong traps are
C$_2$V (\textit{Cf.} Sec. \ref{sec:FIA} ) and we therefore fitted the $E^i_t$ for
SIA
clusters of sizes above $N_{th}$ to 1.4
eV, consistently with the calculations in \cite{anento2013carbon}. For
temperatures above 680 K, we
assume that the C$_2$V complexes begin to dissociate, leaving only single C
atoms in the system, with $E^i_t = 0.6$ eV.

In the case of vacancy clusters, we use the same trapping energies for
vacancies, $E^v_t$, for all temperatures.
For $N^v = $ 1--6, we use DFT values from \cite{becquart2011p60}. For $N^v > 6$,
we assume no trapping, as suggested by \cite{terentyev2011interaction,hepburn2008metallic}.

\section{Simulation of the nanostructure evolution under irradiation}
\label{sec:irradiation}

We simulated an irradiation experiment with the set-up used by Eldrup \textit{et
al}. in \cite{eldrup2002dose}. The irradiation temperature was 343 K and the
simulation box size was $350\times400\times450\times a_0^3$. Periodic boundaries were used in
three dimensions. Cascades were introduced with a rate equivalent to
$7\cdot10^{-7}$ dpa/s. As either SIA or vacancy objects can be exclusively
trapped, two populations of 100 appm traps were employed to include the effect
of C on vacancies and C-V complexes on SIA clusters. The trapping energy depends
on the size of the trapped object, as in Table \ref{SIA_trapping_energies}. Two spherical absorbers for SIA and vacancies,
respectively, were introduced as sinks equivalent to a dislocation density of
$10^{12}$ m$^{-2}$. The grain size was 33 $\mu$m. The simulation was stopped
when 0.734 dpa had accumulated.

\begin{figure}
 \centering
  \includegraphics[width=\columnwidth]{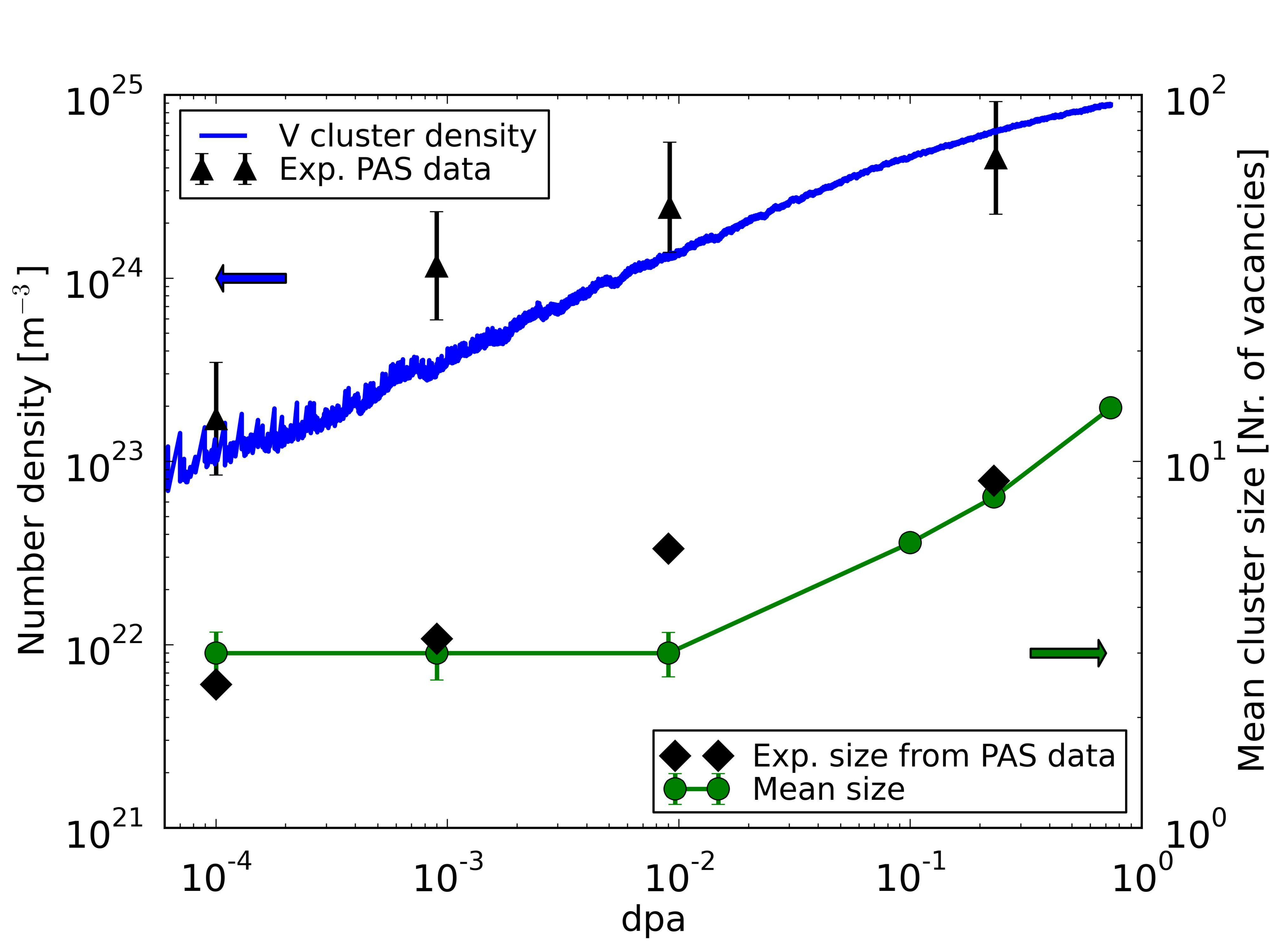}
\caption{The vacancy cluster number density and mean size evolution versus dpa for the best case.
The experimental data are from \cite{eldrup2002dose}.}
\label{R20111221-1.pdf}
\end{figure}

\begin{figure}
 \centering
  \includegraphics[width=\columnwidth]{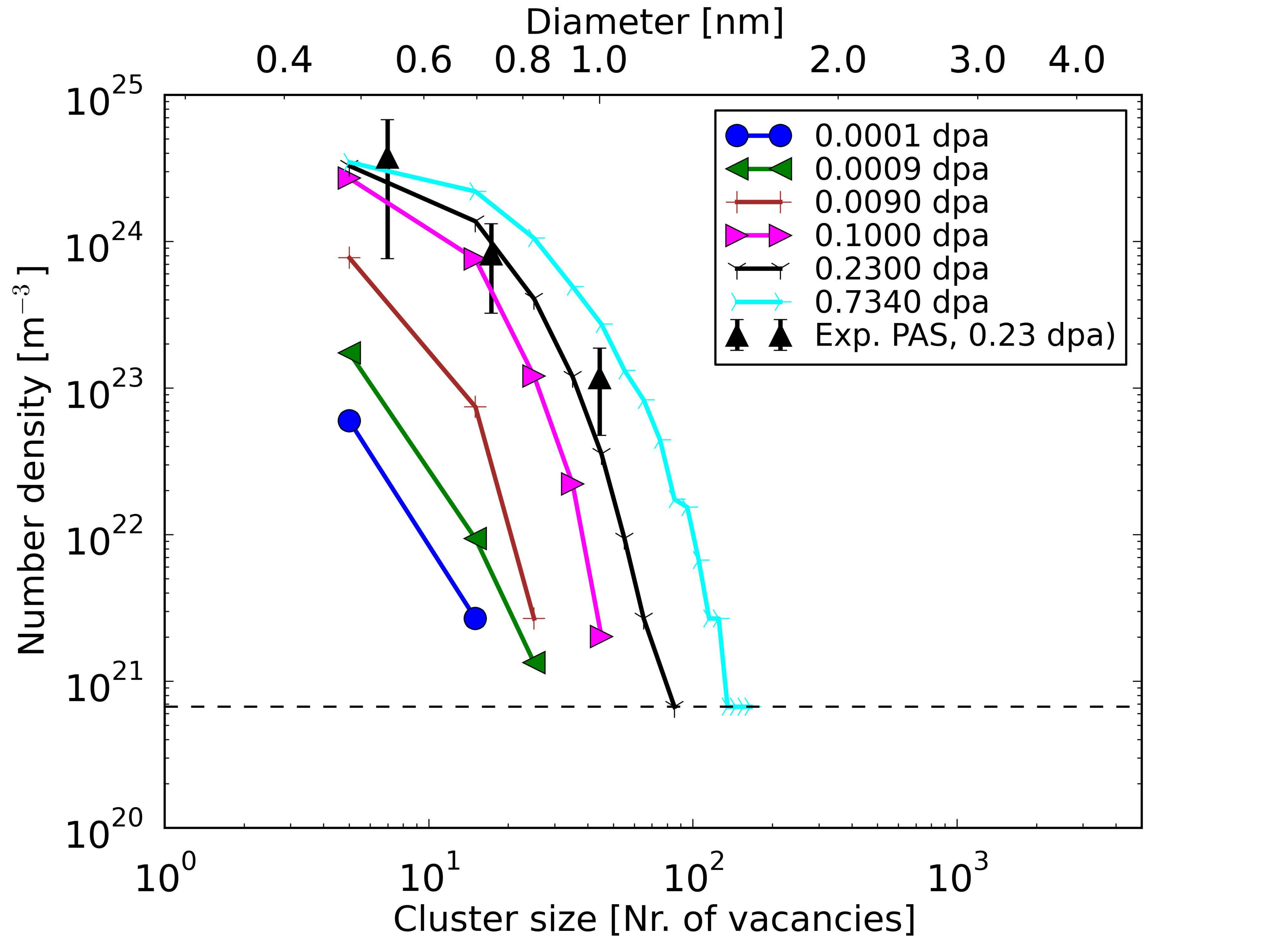}
\caption{The vacancy cluster size distribution for different doses for the best case. The size
increment is by
10 vacancies. The dotted line gives the density for one defect in the simulation
box. The experimental data for 0.23 dpa (triangles) are from
\cite{eldrup2002dose}.}
\label{R20111221-1_vac_binned_size_dist.pdf}
\end{figure}

The vacancy cluster number density and mean size versus dpa, as compared to
experimental data,
are shown in Fig.~\ref{R20111221-1.pdf} for the best case, \textit{i.e.} $N_{th}=29$ (see below for more details). The corresponding size distribution of vacancy
clusters for different doses is shown in Fig. \ref{R20111221-1_vac_binned_size_dist.pdf}.
The results of the simulation show good agreement with the experimental data.

\begin{figure}
 \centering
  \includegraphics[width=\columnwidth]{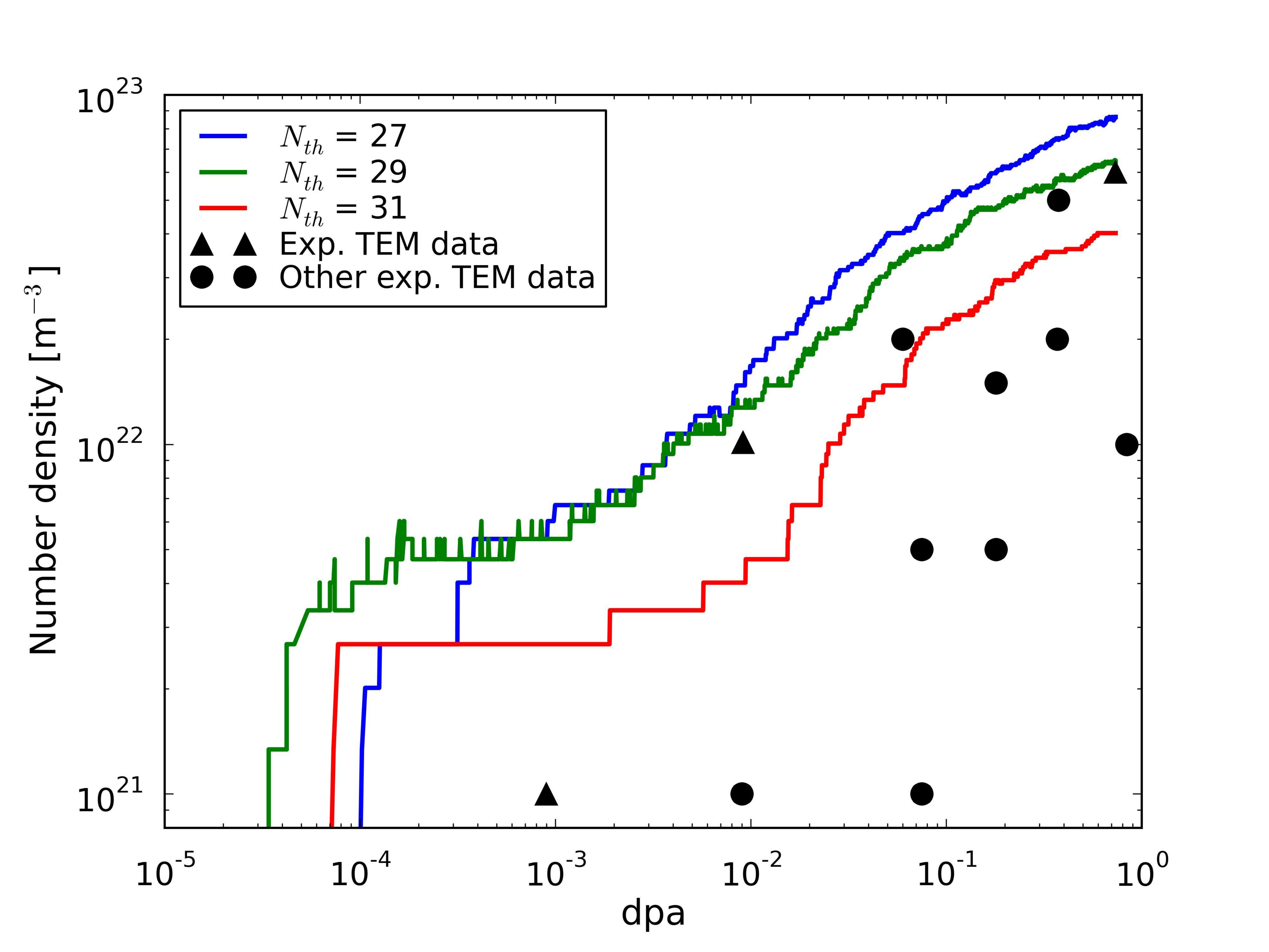}
\caption{Number density of visible SIA versus dpa. $N_{th}=29$ corresponds to our best case. The reference experimental data are denoted with triangles
\cite{zinkle2006microstructure}. Included in the graph are also data from other
comparable irradiation experiments in Fe-C (bullets)
\cite{singh1999effects,eyre1965electron,bryner1966study,robertson1982low,
horton1982tem,takeyama1981}. See \cite{malerba2011review} for full
details.}
\label{R20111221-1_visible_SIA.pdf}
\end{figure}

\begin{figure}
 \centering
  \includegraphics[width=\columnwidth]{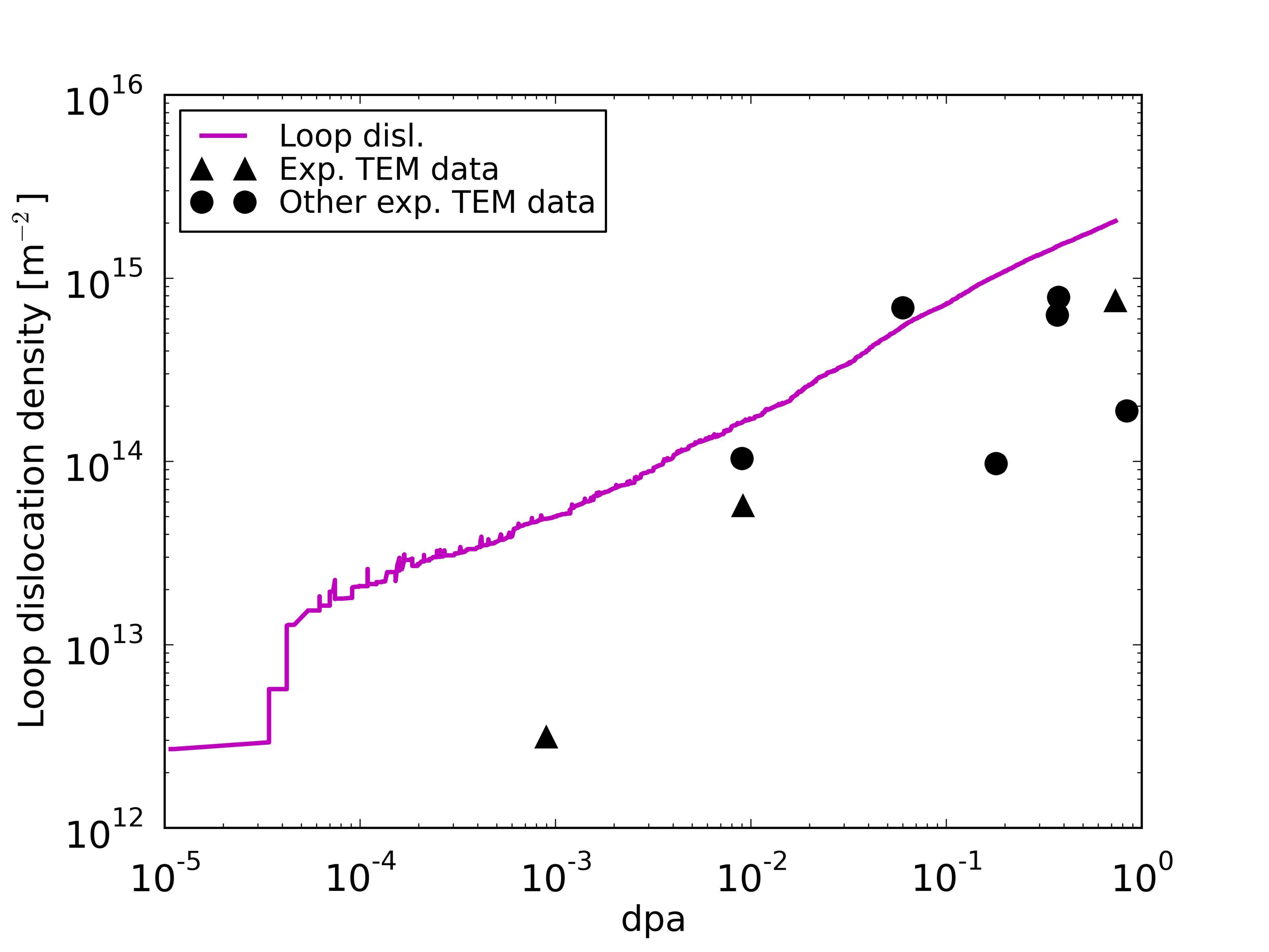}
  \caption{Loop dislocation density versus dpa. The reference experimental
    data are marked with triangles \cite{zinkle2006microstructure}. Included in 
    the graph are also data from other comparable irradiation experiments in Fe-C 
    (bullets)
\cite{singh1999effects,eyre1965electron,robertson1982low,horton1982tem,
takeyama1981}. 
    See \cite{malerba2011review} for further details.}
\label{R20111221-1_loop_disl.pdf}
\end{figure}

The density of visible SIA clusters ($N^i > 50$) versus dose is shown in Fig
\ref{R20111221-1_visible_SIA.pdf}, and compared with a number of experimental
data from TEM examinations of irradiated Fe-C alloys. The threshold, $N_{th}$,
was fitted to the TEM data from Zinkle and Singh, denoted as triangles in Fig.
\ref{R20111221-1_visible_SIA.pdf}. The best fit was obtained, in this case, with $N_{th}=29$. The results for $N_{th}=27$ and 31 are also shown and suggest that a higher $N_{th}$ gives a lower visible SIA cluster density.
Values of $\sim$30 seem indeed reasonable as threshold above which complexes would mainly interact with the centre of the loop and changing the value allows one to remain within the experimental range. Another way of comparing the evolution of SIA clusters with experiments is to
consider the dislocation density due to loops, as shown in Fig.
\ref{R20111221-1_loop_disl.pdf} for $N_{th}=29$. One advantage of this choice is that the
scatter of data from different experiments is significantly reduced. The agreement is good (in both cases) starting from $10^{-3}$--$10^{-2}$ dpa, but the model underestimates by an order of magnitude the dose at which SIA clusters become visible. 

\begin{figure}
 \centering
  \includegraphics[width=\columnwidth]{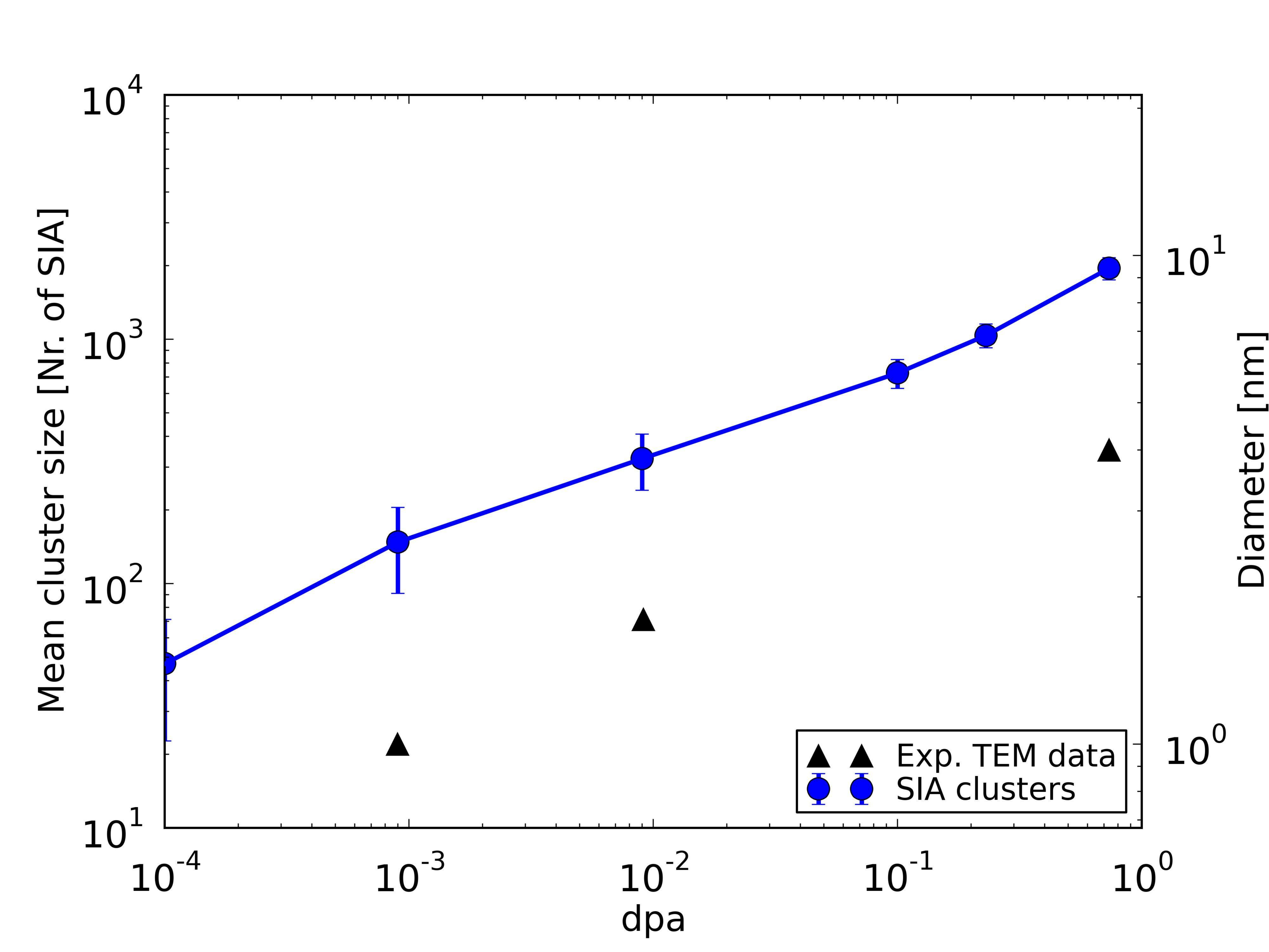}
\caption{The SIA cluster mean sizes and their standard deviations for different
dpa. The experimental data are from \cite{zinkle2006microstructure}.}
\label{R20111221-1_SIA_mean_size.pdf}
\end{figure}

\begin{figure}
 \centering
  \includegraphics[width=\columnwidth]{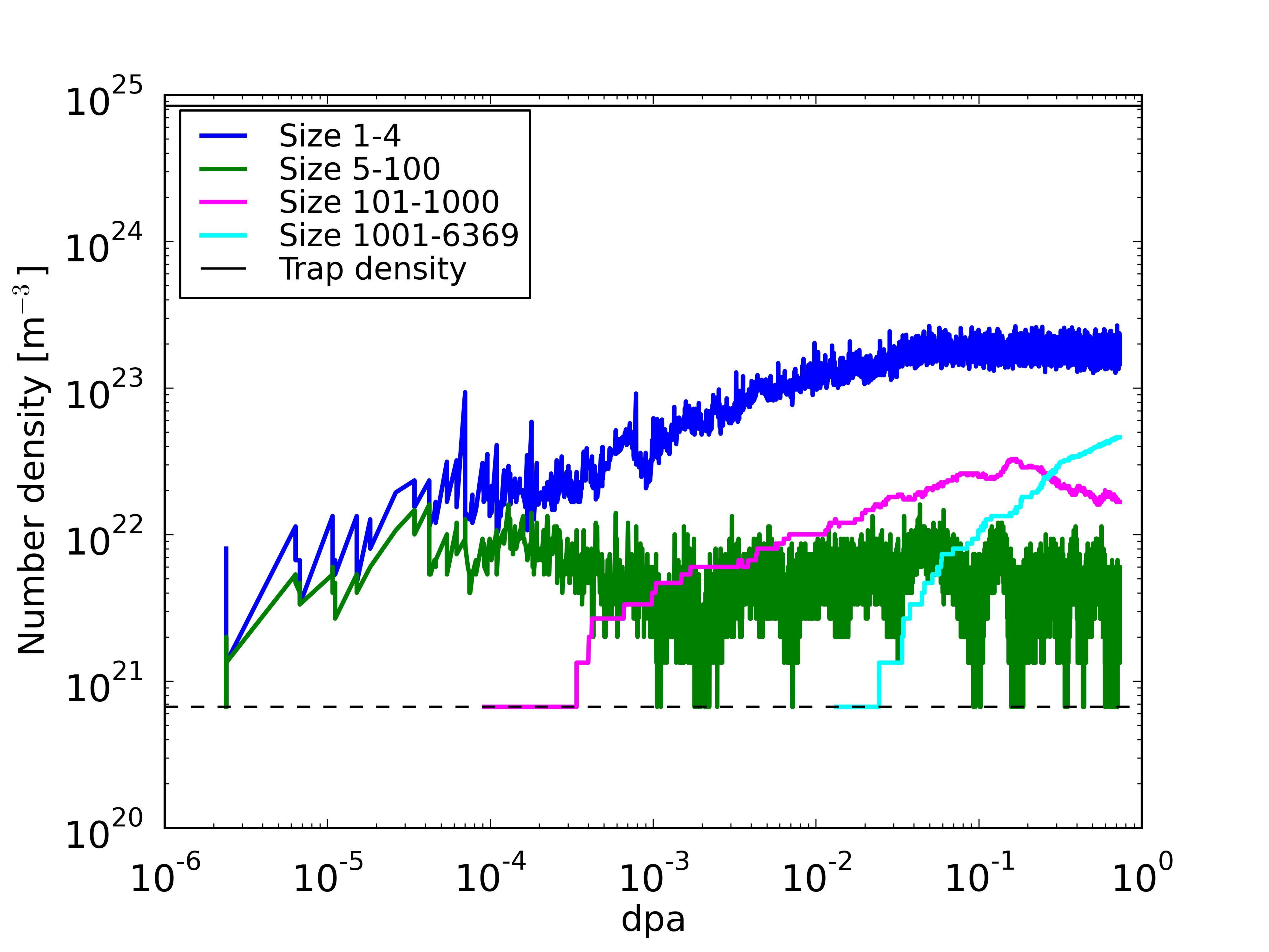}
\caption{The SIA cluster size distribution evolution as function of dpa. The
dotted line gives the density for one defect in the simulation box.}
\label{R20111221-1_SIA_size_evolution.pdf}
\end{figure}

The mean cluster size at different doses is shown in Fig.
\ref{R20111221-1_SIA_mean_size.pdf}. The mean sizes are overestimated compared to the reported experimental data \cite{zinkle2006microstructure}, although the trend is correct. The evolution of the size distribution of SIA
clusters versus dpa is shown in Fig.~\ref{R20111221-1_SIA_size_evolution.pdf}.

\section{Simulation of the nanostructure evolution during isochronal
annealing}\label{sec:annealing}

We simulated a complete post-irradiation-annealing experiment in a box
of size $350\times400\times450\times a_0^3$. In the reference experiment \cite{eyre1965electron}, iron was irradiated at 333
K to 0.37 dpa at a dpa rate of $5\cdot10^{-8}$
dpa/s. Similarly to the previous case, we introduced 100 appm
traps for SIA objects, as well as 100 appm
traps for vacancy objects. The traps, that we assume are representing C and CV
and CV$_2$ at low temperatures, had the same defect-size-dependent and
temperature-dependent-binding energies as described in Sec. \ref{sec:trap_parameters} and in Table \ref{SIA_trapping_energies}. No values for the dislocation density and grain sizes are reported in
\cite{eyre1965electron}. We used the same values as in
the previous simulation in Sec. \ref{sec:irradiation}

\begin{figure}
\centering
\includegraphics[width=\columnwidth]{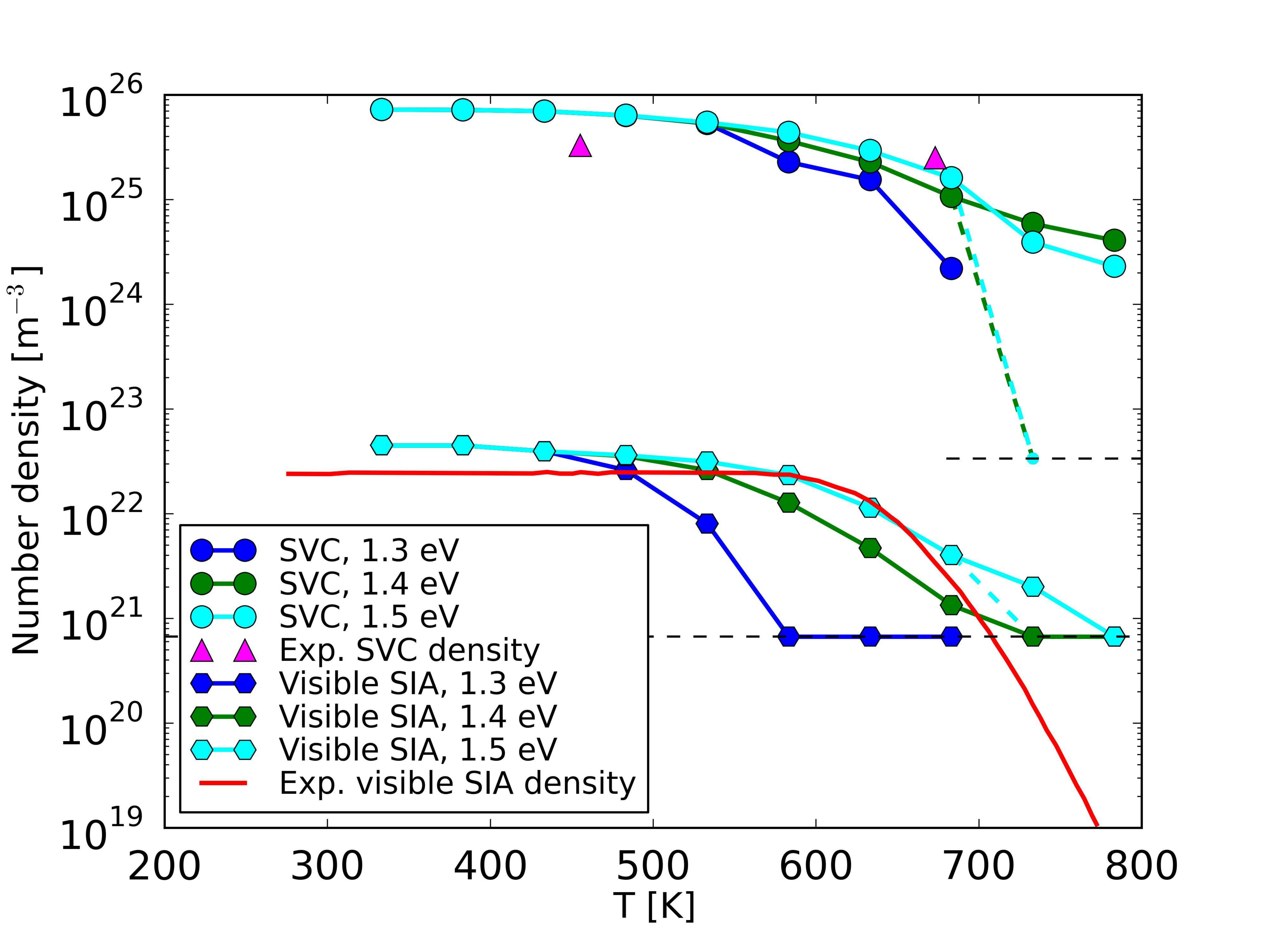}
\caption{The density of visible SIA clusters and SIA in visible clusters (SVC) as a
function of temperature during the simulated isochronal annealing. Different trapping energy for SIA, $E^i_t =$ 1.2--1.5 eV, are used at the temperature of 483 K and higher. Above 733 K, if the trapping energy was decreased to $E^i_t= 0.6$ eV, all clusters
disappear, as indicated by the dashed coloured lines. The solid line and
the triangular dots are the experimental visible SIA cluster density and the SVC density, respectively, from \cite{eyre1965electron}. The latter points are calculated from the reported visible SIA densities and the cluster diameter data. The dotted black lines indicates the density
corresponding to one object in the box (lower line) and the minimum SVC simulation resolution (higher line), respectively.}
\label{R20120518-00_2.pdf}
\end{figure}

After irradiation the system underwent isochronal annealing, where the
temperature was increased by 50 K after every hour (3600 s), until a final
temperature of 733 K was reached. We simulated the evolution with temperature and traced the number
density of visible SIA clusters, as well as the number of SIA in visible
clusters (SVC), as shown
in Fig. \ref{R20120518-00_2.pdf} to compare with corresponding
experimental data.

Before the temperature was raised to 483 K,
we assumed that carbon atoms should have become mobile and thus able to form the
stronger C$_2$V traps, as is suggested by the study performed in Sec.
\ref{sec:CV_results} (Fig. \ref{R20111102-1.pdf}).
We thus raised the trapping energy for large SIA (larger than $N_{th}$) to different values, $E_t^i= 1.3$, 1.4 and 1.5 eV. According to MD simulations the binding energy between a SIA cluster and a C$_2$V complex is 1.4--1.5 eV \cite{anento2013carbon}. Our results indeed suggest a value between 1.4 and 1.5 eV for $E_t^i$. The trapping energies of the vacancy traps were unchanged, but their number was correspondingly reduced by $50 \%$.

Using the \textit{ab initio} values for the C$_2$V complex from Table
\ref{FV_parameters_table},
the dissociation energy can be calculated as $E_{diss} = M^{fv}_f+B^{fv}_f =
1.90$ eV, which, using the formula for thermally activated events,
\begin{equation} 
\Gamma_e = \nu^{fv} \exp{\left(\frac{-E_{diss}}{k_BT}\right)}
\end{equation} 
gives an average lifetime of the complex, $\Gamma_e^{-1} = 18.4$ s, at $T =
683$ K, rendering the complex unstable. Above this temperature, we can thus
assume only C atoms to remain in the box, all C-V complexes being dissolved.
Thus, before the temperature was raised to 733 K, we lowered the trapping energy
for large SIA clusters to 0.6 eV, \textit{i.e.} the binding energy between C and
large SIA clusters \cite{terentyev2011interaction}. The number of traps were again raised by 50 \% to the original density.
Due to the lowered trapping energy all SIA and vacancy clusters disappear, mostly due to recombination (in the absence of other efficient sinks). This is also seen in the experimental data. If the trapping energy is not lowered, with $E_t^i=1.4$ or 1.5 eV, some SIA clusters will remain longer in the box than the experimental data show. With $E_t^i<1.4$ eV the clusters disappeared too fast.

We also did a separate study where we kept the pre-irradiated system at the constant room temperature of 290 K to see if any changes could be expected from normal storage of the irradiated material. While the vacancy and SIA densities dropped a bit and the mean cluster sizes rose a bit, no significant changes could be observed after a simulation time of about one year.

\section{Discussion}\label{sec:discussion}

\subsection{Small C-V complexes}
  
The study of small C-V complexes in Sec. \ref{sec:FIA} has inherent limitations as
the formation of large V and C-V complexes is deliberately forbidden, while SIA are
altogether absent. Still, as steady-state is reached
after a relatively short time and $\sim$10$^{-4}$ dpa, it can be assumed that
the same steady-state would be reached very quickly in a real irradiation, before
the SIA cluster population starts to build up. In this sense, the use of traps
instead of the more cumbersome explicit treatment of all possible C-V clusters
is acceptable. 

We used two regimes, where one was under-saturated and the other was over-saturated with vacancies. The second one   might look closer to a real irradiation but in reality whether one regime or the other is correct it will depend on the actual C content in the matrix and on the efficiency of the sinks. The two regimes can be considered as representative limiting cases. Concerning the density of CV$_2$, both regimes are characterised by more or less the same behaviour. The main difference is that in the under-saturated regime, there is an abundance of free C, whereas in the over-saturated regime there is not, as they are all bound to CV$_2$ after the irradiation.
In Sec. \ref{sec:CV_results}, we clearly see that CV$_2$ is the dominating
complex at the irradiation temperature chosen in this work, especially if we disregard the
weakly trapping C atoms. This is logical as C atoms at low temperatures are
virtually immobile and V have more time to bind to C defects.

Based on the parameters deduced from \textit{ab initio} data, above 480 K, C is mobile
and will join with the CV$_2$ complexes, creating the C$_2$V complexes, as the second V
becomes weakly bound in the C$_2$V$_2$ complexes and is quickly emitted. As there can be a difference of 0.6--1.0 eV for the binding energy of a C$_2$V and a CV$_2$ complex to the same point of a SIA cluster (centre or edge) \cite{anento2013carbon}, there are clearly two different regimes for SIA traps at lower and higher temperatures, respectively.

\subsection{Cluster evolution under irradiation}

Sec. \ref{sec:irradiation} demonstrates that our model correctly reproduces the
vacancy density cluster evolution under irradiation at 343 K. For the SIA
clusters, we compare the simulation data in two ways with the available
experimental data: in terms of density of visible loops, and in terms of
loop dislocation density, as functions of dpa. As all the experimental data are
retrieved from TEM studies, which we expect to have fairly large error bars, up
to about one order of magnitude, we can say that we have good agreement also for
the SIA evolutions, comparing specifically to the data by Zinkle and Singh
\cite{zinkle2006microstructure}, to which the visible SIA density was in fact
fitted by varying the threshold parameter, $N_{th}$, (the only
fitting parameter used here). The other TEM data show a rather extensive
scatter, illustrating the large experimental uncertainties. It can be noted that
the newer experiments (our reference experiment is the most recent one) tend to have higher densities than the older ones, which might be explained by the higher resolution of newer electron microscopes. 

The SIA mean cluster sizes are systematically larger, compare to the experimental data by S.J. Zinkle and B.N. Singh \cite{zinkle2006microstructure}, but their reported mean sizes are on the other hand surprisingly small, down to $\sim$1 nm in diameter, (they report a TEM resolution of 0.5 nm, which is a significantly smaller than the more standard value of 1.5 nm, which we indeed have used as a lower size limit for visible clusters). A lower value of the limit for visibility in the analysis for the simulation data does not improve our results, though the experimental data might thus in fact be a bit underestimated in case of the mean size. 

The defect size dependence of the trapping energy to the SIA traps proved
crucial for the model. With a constant trapping energy for all sizes,
all SIA clusters would eventually merge together to a single large cluster,
at high enough dpa. This would happen independently of the simulation box
size, which is not physical. The importance of a size dependence of the emission
rate from traps was already evidenced in \cite{lee2009kinetic}, however here we
provide a physical interpretation for it, in terms of changing dominant
mechanism in the interaction between C-V complexes and SIA clusters (edge versus
central interaction).

\subsection{Post-irradiation annealing}

To be successful, the simulation of the isochronal annealing in Sec.
\ref{sec:annealing}, had to be divided into three different temperature ranges where
the trapping energy changed according to the results from Sec. \ref{sec:FIA}. In
particular, the trapping energy, $E^i_t$, for large SIA clusters above the
threshold size, $N_{th}$ was changed from the first to the second stage to agree
with the values attributable to the dominant C-V cluster species,
\textit{i.e.} from $E^i_t \sim 1.2$ eV for CV$_2$ to 1.4 eV for C$_2$V. 1.2 eV is enough to completely trap SIA clusters at the first stage, but not enough at the higher temperatures at the second stage ($>$483 K), why the stronger value of 1.4 eV of the C$_2$V \cite{anento2013carbon} complex is necessary. Indeed, the MD study by N. Anento and A. Serra suggests that the binding energy of CV$_2$ should be 1.4 eV as well \cite{anento2013carbon}, which is still in good agreement with our results, as a higher trapping energy at the first stage does not change anything.

In the third stage, the weak traps, corresponding to isolated C atoms,  interact
very little with the large remaining SIA clusters, but still prevent them to
all cluster together into a single cluster. In the end all clusters either
recombined or were absorbed by the grain boundaries or other sinks. 

The overall temperature evolution shows fair agreement for all three stages of
the annealing with the experimental data for both the visible SIA density, as
well as SIA in visible clusters.

\subsection{General discussion}

Clearly, the present model has limitations related to the fact that the
physics of the formation of complexes between C atoms and point-defects,
vacancies but also SIA clusters, is implicit in the parameterization of the
trapping energies. A more evolved model should include explicitly all the
ingredients and provide spontaneously all the effects. Yet, it is believed that
the present work represents a significant step forward towards the development
of nanostructural models for Fe alloys, including steels, by providing a
reference for the elaboration of more complete ones.

An issue that was not addressed here concerns the sensitivity of the model to
the variation of key parameters on which uncertainties remain, first of all the
SIA cluster threshold size explicitly used to fit the results. Moreover, it is
of high interest to perform a wide exploration on the effect of irradiation
parameters such as flux. These studies have actually been performed; however,
because of the extensive amount of results they led to, it has been preferred to
report them in a separate paper \cite{jansson2013okmc}.

\section{Conclusions}\label{sec:conclusion}

We have presented an OKMC model based on the use of a physically fully
motivated parameter set, which succeeds in describing the nanostructure evolution
in  
Fe-C systems  under irradiation at 333--343 K ($<$100\textdegree C) and during
post-irradiation annealing. 

Key for the success of the model is the deep understanding, based on \textit{ab
initio} and interatomic potential calculations, of the interaction between C
atoms and point-defects, to form stable complexes with vacancies that are in
turn capable of binding migrating SIA clusters. In this paper the physics is
simplified by introducing appropriately parameterized traps for point-defect
clusters, following the example of previous models. Next step will be the
implementation of a full set of parameters treating explicitly the formation of
complexes involving C.

\section*{Acknowledgement}

This work was carried out as part of the PERFORM60 project of the 7th Euratom Framework Programme, partially supported by the European Commission, Grant agreement number FP7-232612. The authors wish to thank N. Anento, C. Becquart, A. De Baker, C. Domain, A. Serra, and D. Terentyev for advice, assistance and fruitful discussions during the performance of this work.



%% References


\bibliographystyle{model1a-num-names}
%\bibliographystyle{apsrev} % The article titles are removed.
% \bibliography{vjansson,vjansson_publications}
\bibliography{vjansson,vjansson_publications}






% \end{thebibliography}

\end{document}